\documentclass{article}

\usepackage{arxiv}

\usepackage[utf8]{inputenc}
\usepackage[T1]{fontenc}
\usepackage{hyperref}
\usepackage{url}
\usepackage{booktabs}
\usepackage{amsfonts,amsmath,amsthm,amssymb}
\usepackage{nicefrac}
\usepackage{graphicx}
\graphicspath{ {./images/} }
\usepackage{color}
\usepackage[english]{babel}
\usepackage{amsmath,amssymb,amsthm}
\usepackage{multirow}
\usepackage{ctable}
\usepackage[normalem]{ulem}
\usepackage{float}
\usepackage{placeins}
\usepackage{mathrsfs}
\usepackage{xcolor}
\usepackage{textcomp}
\usepackage{manyfoot}
\usepackage{algorithm}
\usepackage{algorithmicx}
\usepackage{algpseudocode}
\usepackage{listings}
\usepackage{anyfontsize}
\usepackage{subcaption}
\usepackage{bm}

\DeclareMathOperator*{\sgn}{sgn}

\title{Localization region detection with directionality estimation in a two-dimensional hexagonal crystal lattice model\thanks{Preprint.}}

\author{
 Filips Kozirevs \\
 Faculty of Science and Technology \\
 University of Latvia \\
 Jelgavas Street 3\\
 Riga, LV-1004, Latvia \\
 \texttt{filips.kozirevs@lu.lv}
 \And
 J\={a}nis Baj\={a}rs \\
 Faculty of Science and Technology \\
 University of Latvia \\
 Jelgavas Street 3\\
 Riga, LV-1004, Latvia \\
 \texttt{janis.bajars@lu.lv}  
}

\begin{document}

\maketitle

\begin{abstract}
This work is devoted to data-driven identification of discrete breathers in numerical simulations of a two-dimensional crystal lattice using locally sampled wave data. Different lattice wave datasets are considered, with data collected from regions of different shapes and sizes defined by the lattice particles in mechanical equilibrium. Specifically, in addition to regions with a regular hexagonal shape, one- and quasi-one-dimensional regions reflecting the quasi-one-dimensionality of discrete breathers in two-dimensional hexagonal crystal lattices are proposed. To improve numerical efficiency, dataset dimensionality is reduced using Principal Component Analysis, and highly accurate Support Vector Machine classifiers are trained to distinguish between linear and nonlinear wave data. The obtained classifiers, together with the sliding window method, are applied to detect localization regions in two-dimensional hexagonal crystal lattice numerical simulations. High-precision algorithms for detected localization region segmentation and localized wave directionality estimation within the detected regions are further proposed, and their performance is evaluated. The presented methods are successfully applied to detect localized waves and their collision regions, as well as their directionality, performing a numerical study of stationary and traveling two-dimensional discrete breather interactions. Qualitatively better results are obtained when considering wave-data collection regions respecting the quasi-one-dimensional nature of two-dimensional discrete breathers in the hexagonal crystal lattice model.
\end{abstract}


\section{Introduction}

Computational powers available to researchers are constantly increasing, and, therefore, performing extensive numerical simulations as well as storing simulation data and analyzing it has become more feasible nowadays. This has given rise to rapid development of data-driven methods for studying physical systems~\cite{montans2019data}. Such data mining of experimental or numerical simulation data allows for exploring and discovering new insights into complex dynamical systems, as well as helps with automating data processing. One of the research directions where data-driven methods have recently found their application is the identification of \emph{intrinsic localized modes} (ILMs) also known as \emph{discrete breathers} (DBs)~\cite{bajars2022data, dogkas2022identifying, Tsironis2025}, which are localized time-periodic crystal lattice waves with internal oscillation~\cite{flach2008discrete}. 

Discrete breathers is an area of active theoretical, numerical, and experimental research. Existence of stationary DBs, that are time-periodic but non-traveling localized modes, in crystal lattice models has been proven~\cite{mackay1994proof, aubry2006discrete}, while existence of traveling DBs remains unestablished~\cite{flach2008discrete, flach1999moving}. Nevertheless, performing numerical experiments, the phenomenon of ILMs is observed in different crystal lattice models, and it is possible to generate stationary as well as traveling DBs in simulations~\cite{bajars2022data, marin1998,  marin2001breathers, dou2011breathers, bajars2015nonlinear, bajars2021two, duran2022moving, abdullina2025excitation}, to name but a few. In addition to excited DB solutions, which are near-periodic localized wave solutions immersed in a noisy background or thermal bath, considered when their lifetime, studied in~\cite{archilla2025thermal}, is of particular importance, the so-called numerically exact to machine precision DB solutions can be computed~\cite{marinaubry96, Pterobreathers, BajarsArchilla2D} to investigate their existence as well as spectral and stability properties.   

The attention of researchers is attracted to DBs because the phenomenon of \emph{hyperconductivity} is experimentally confirmed, meaning that such moving localized waves can transfer charge in silicates~\cite{Russell_2017, russell2019hyperconductivity, russell2020localized}, which is thought to be attributed to the formation of long dark lines in muscovite mica crystals~\cite{archilla2015quodons}. Moreover, DBs have been found present in models of DNA and proteins, as well as in Josephson arrays~\cite{Fakhretdinov2013, DNAchain, nicolai2015intrinsic, piazza2008discrete, Mazo03}, to name but a few. 

Such a phenomenon when an extra charge or hole is bound and can hop along ion chains in silicates has been theoretically modeled and numerically investigated in~\cite{Archilla_2024}, in addition to conceptually related works~\cite{archilla2015quodons, kalosakas1998, cuevas2019, Ros2011, chetverikov-epjst2013}, to name but a few. Numerically exact solutions of stationary and moving DBs coupled with an extra charge or hole have been studied for a phenomenological model in~\cite{AxiomsJBJA}, where spectral properties of such time-periodic solutions, which are also called \emph{polarobreathers}, have been investigated. It was shown that DBs can carry an extra charge for long distances.

Given the vast interest in DBs, the abundance of numerical data, and recent advances in data-driven methods, the analysis of crystal lattice simulation data is a promising approach for studying ILMs and their properties. Machine learning algorithms were considered in~\cite{bajars2022data, dogkas2022identifying, Tsironis2025} for the identification and classification of DBs in crystal lattice simulations. The main objective of the present work is to generalize and extend the approach proposed in~\cite{bajars2022data} to the case of a two-dimensional (2D) crystal lattice. The data-driven methodology presented in~\cite{bajars2022data} relies on collecting locally linear and nonlinear wave data, which is then used to train classifiers to identify localization regions in numerical simulations of crystal lattice dynamics when combined with the sliding window method, scanning the simulation data along the whole lattice. While the current approach is based on locally sampled dynamics data, in~\cite{dogkas2022identifying, Tsironis2025}, the authors considered distinguishing DBs from phonons by using Convolutional Neural Networks trained on 2D images of lattice wave time evolutions. Their approach also allows for determining the underlying nonlinear on-site potentials that have generated DBs. The advantage of the currently proposed methodology is that it is local in space, does not rely on time series data, and provides an automated means for localization region detection with further segmentation. In addition, the current approach has different benefits compared to global Fourier, Gábor, and wavelet transforms~\cite{Hori93, Forinash98, Annise19}, which provide valuable spectral information and its evolution in time or space but are limited in determining the actual localization sites at a given time.

To extend the proposed methodology of~\cite{bajars2022data}, which was entirely developed for one-dimensional crystal lattice models, to higher crystal lattice dimensions, further steps are taken. That is, in this work, highly accurate data-driven algorithms relying only on locally sampled data at a specific time are described and applied for localized wave classification and localization region identification in 2D crystal lattice numerical simulations, also taking into account the localized wave geometrical properties. Moreover, a data-driven method of high precision is described and applied for localized wave directionality estimation in crystal lattice simulations based on the identified localization regions. The performance of all the mentioned methods is evaluated and compared, demonstrating that not only can the methodology be applied in higher crystal lattice dimensions, but also directionality of localization can be estimated, which may be important when DBs change their propagation direction after collisions. It is also shown that qualitatively better results are obtained when geometrical properties of DB solutions are taken into account. Thus, the proposed data-driven methods can stimulate further automated post-processing of numerical simulation data and gaining a better understanding of DB properties and energy localization in nonlinear dynamical systems.

The paper is organized in the following way. In Section~\ref{sec:mathematical_model}, the mathematical model of the 2D crystal lattice considered in this work is introduced. The process of localized wave data generation and collection is described in Section~\ref{sec:data_collection}. Section~\ref{sec:classification} is devoted to the description of the lattice wave data dimensionality reduction and classification. The dimensionality reduction and classification algorithms are applied, and their performance is evaluated. In Section~\ref{sec:numerical_results}, the previously trained classifiers are applied in crystal lattice numerical simulations for finding localized wave regions based on locally sampled data. Then, an algorithm for localization region segmentation and localized wave directionality estimation is presented, and its performance is analyzed. A simulation study of DB interactions for evaluating the precision of the considered methods in cases when many DBs of different types propagate and interact is described at the end of Section~\ref{sec:numerical_results}. Two DB collisions are studied in detail. In Section~\ref{sec:conclusions}, conclusions are provided. The process of training the classifiers used in this work is comprehensively described in Appendix~\ref{ap:SVC_training}.

\section{Mathematical model}\label{sec:mathematical_model}

Without loss of generality, in this work, a 2D hexagonal crystal lattice model is considered. This model was proposed in~\cite{bajars2015nonlinear} to model a 2D K–K sheet layer of muscovite mica crystal and has been further explored in~\cite{BajarsArchilla2D,bajars2021two}, studying frequency–momentum representation of traveling breathers and their scattering after collisions, respectively. The model considered below can also be viewed as a phenomenological model for studying 2D DBs, owing to the ease of generating such localized wave solutions.

Closely following~\cite{bajars2015nonlinear}, the $N$ particles of the crystal layer are arranged in a hexagonal lattice, their dynamics is Hamiltonian, and periodic boundary conditions are applied. The corresponding dimensionless Hamiltonian function is
\begin{equation}\label{eq:Hamiltonian}
H=\mathrm{K}+\mathrm{U}+\mathrm{V}=\sum_{i=1}^N\left(\frac{1}{2}\|\dot{\mathbf{q}}_n\|^2+U(\mathbf{q}_n)+\frac{1}{2}\sum_{l\in\mathcal{L}_n}V(r_{n,l})\right),
\end{equation}
where $\mathrm{K}$ is the total kinetic energy, $\mathrm{U}$ is the total on-site potential energy used to model the interaction of the layer particles with the crystal layers above and below it, $\mathrm{V}$ is the total particle interaction potential energy, $\mathbf{q}_n=(q_{nx},\,q_{ny})^T\in\mathbb{R}^2$ is the position of the $n$-th particle and its time-derivative $\dot{\mathbf{q}}_n=\mathbf{p}_n=(p_{nx},\,p_{ny})^T\in\mathbb{R}^2$ is the corresponding momentum, $\mathcal{L}_n$ is the set containing indices of the nearest neighbors of the $n$-th particle and $r_{n,l}$ is the Euclidean distance between the $n$-th and the $l$-th particles taking into account the periodic boundary conditions. The on-site potential is a smooth periodic function
\[
U(\mathbf{q}_n)=\frac{2}{3}\left(1-\frac{1}{3}\left(\cos{\left(\frac{4\pi q_{ny}}{\sqrt{3}}\right)}+\cos{\left(\frac{2\pi(\sqrt{3}q_{nx}-q_{ny})}{\sqrt{3}}\right)}+\cos{\left(\frac{2\pi(\sqrt{3}q_{nx}+q_{ny})}{\sqrt{3}}\right)}\right)\right)
\]
with hexagonal symmetry~\cite{yang2011rectification}. To model the particle interactions, the Lennard-Jones potential
\[
V(r_{n,l})=\varepsilon\left(\left(\frac{1}{r_{n,l}}\right)^{12}-2\left(\frac{1}{r_{n,l}}\right)^{6}\right)
\]
is considered, where the parameter $\varepsilon>0$ describes the potential well depth ratio of the potentials $V$ and $U$, respectively~\cite{adams2001bonding}. As in~\cite{bajars2015nonlinear,bajars2021two,BajarsArchilla2D}, it is considered that $\varepsilon=0.05$. The proposed model is a generalization of the one-dimensional (1D) crystal lattice model considered in~\cite{bajars2022data}. As in the 1D case, only the interactions between particles and their nearest neighbors are considered. However, in the 2D lattice, each particle has six nearest neighbors, whereas in the 1D lattice -- only two. Moreover, in the 2D case, there are three crystallographic directions characterized by the following direction cosine vectors:
\begin{equation}\label{eq:directions}
(1,\,0)^T,\quad (1/2,\,\sqrt{3}/2)^T\quad\text{and}\quad(1/2,\,-\sqrt{3}/2)^T.
\end{equation}

From the equation~\eqref{eq:Hamiltonian}, the Hamiltonian equations can be obtained:
\begin{equation}\label{eq:Hamiltonian_system}
\begin{aligned}
\dot{\mathbf{q}}_n ={} & \nabla_{\mathbf{p}_n}H=\mathbf{p}_n, \\
\dot{\mathbf{p}}_n ={} & -\nabla_{\mathbf{q}_n}H=-\nabla U(\mathbf{q}_n)-\frac{1}{2}\sum_{l\in\mathcal{L}_n}\nabla_{\mathbf{q}_n}V(r_{n,\,{l}}),
\end{aligned}
\end{equation}
which are further solved numerically using the second-order Verlet method with the time step $\tau=0.01$~\cite{hairer2003geometric}. One of the most significant properties of the Verlet method is its simplecticity, which ensures that values of the numerically obtained Hamiltonian function are close to its initial value for a long period of time, and the absolute Hamiltonian error is of order two. Moreover, the Verlet method is time-reversible.

Significantly, the proposed crystal lattice model exhibits ILM solutions, which are also known as DBs. In the 2D lattice case, DBs are quasi-one dimensional (quasi-1D), which means that the energy of a DB is mostly concentrated only on one layer of crystal lattice particles having one of the crystallographic directions~\eqref{eq:directions}. This crystallographic direction is further called the \emph{direction of the DB}~\cite{bajars2015nonlinear}.

\section{Wave-data generation and collection}\label{sec:data_collection}

In this work, a data-driven approach for localized wave region detection and directionality estimation is considered, which means that the proposed methods are based on the simulation data. This section is devoted to the description of the process of creating the datasets containing locally sampled lattice wave data, which are further used for training the classifiers, described in Section~\ref{sec:classification}. All calculations in this work are performed using the programming language \emph{Python} together with such libraries as \emph{NumPy} and \emph{scikit-learn} as well as the programming language \emph{C}~\cite{harris2020array, pedregosa2011scikit-learn}.

In order to generate a DB solution with a specific direction in a numerical simulation, a layer of particles with the appropriate direction has to be chosen, together with a region of three or four particles in this layer, depending on the wave type, where the DB will be initialized. The initial conditions of a special form are applied, numerically solving the equations~\eqref{eq:Hamiltonian_system}:
\begin{itemize}
\item for generating a stationary DB,
\[
\begin{aligned}
\mathbf{p}^{0} ={} & \gamma (p^0_1,\,\ldots,\,p^0_{n_0-1},\,-1,\,2,\,-2,\, 1,\,p^0_{n_0+4},\,\ldots,\,p^0_N)^T\in\mathbb{R}^N, \\
\mathbf{q}^{0} ={} & (0,\,\ldots,\,0)^T\in\mathbb{R}^N,
\end{aligned}
\]
where
\[
p^0_n=0,\quad\text{if}\quad n\in\{1,\,\ldots\,,\,n_0-1,\,n_0+4,\,\ldots,\,N\},\quad\text{and}\quad\gamma>0;
\]
\item for generating a traveling DB,
\[
\begin{aligned}
\mathbf{p}^{0} ={} & \gamma (p^0_1,\,\ldots,\,p^0_{n_0-1},\,-1,\,2,\, -1,\,p^0_{n_0+3},\,\ldots,\,p^0_N)^T\in\mathbb{R}^N, \\
\mathbf{q}^{0} ={} & (0,\,\ldots,\,0)^T\in\mathbb{R}^N,
\end{aligned}
\]
where
\[
p^0_n=0,\quad\text{if}\quad n\in\{1,\,\ldots,\,n_0-1,\,n_0+3,\,\ldots,\,N\},\quad\text{and}\quad\gamma\in\mathbb{R}\setminus\{0\}.
\]
\end{itemize}
Here, the absolute value of $\gamma$ characterizes the generated wave amplitude and energy, see~\cite{bajars2015nonlinear}. Furthermore, in the case of a traveling wave, the sign of $\gamma$ indicates the direction of the wave motion (not to be confused with the DB direction). To generate linear phonon waves, the components of the initial displacements and momenta of the lattice particles are drawn from the uniform distribution $\mathcal{U}(-0.01,\,0.01)$. It is important to note that these initial conditions do not produce exactly time-periodic stationary or traveling DB solutions of the whole system, but rather DB solutions immersed in a weakly noisy background, depending on the constant $\gamma$.

\begin{figure}[t!]

\begin{subfigure}{\textwidth}
\hspace{6em}\includegraphics[width=0.77\textwidth]{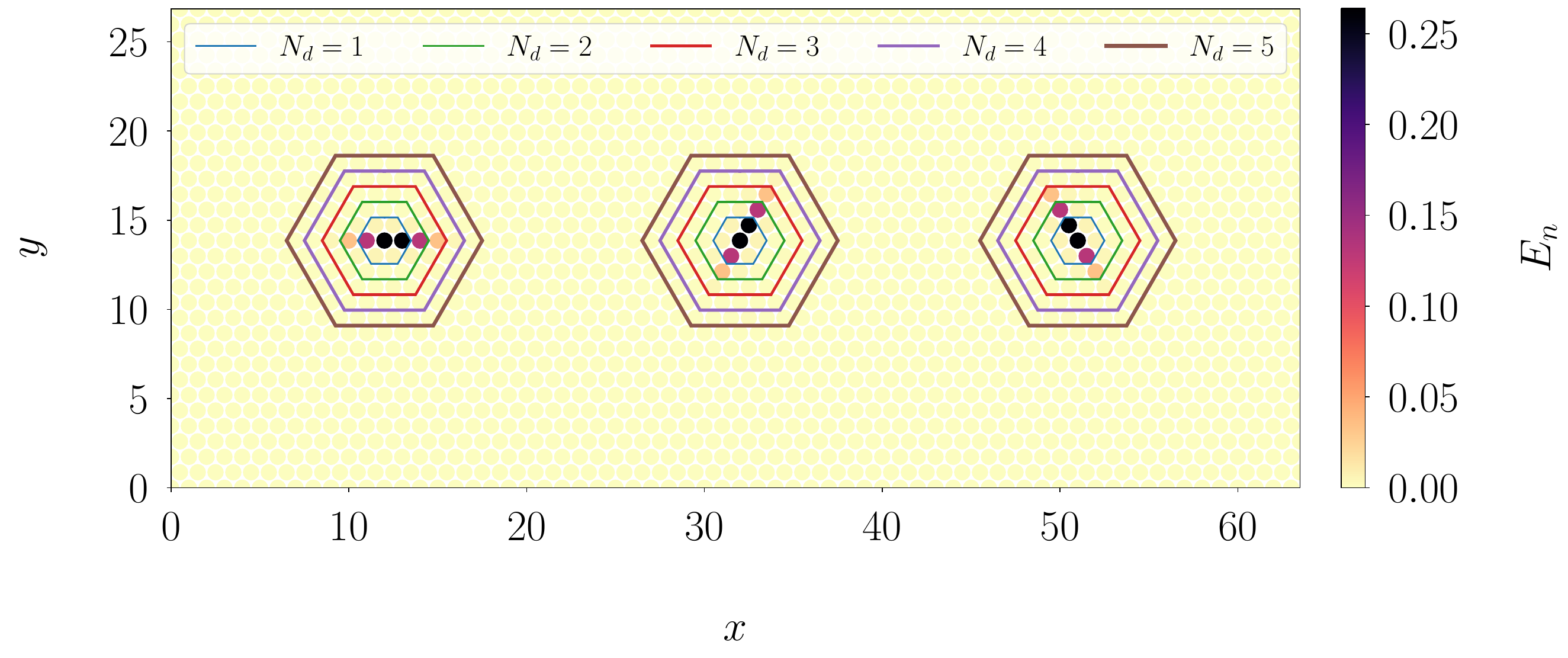}
\caption{2D hexagonal regions}
\end{subfigure}

\bigskip

\begin{subfigure}{\textwidth}
\hspace{6em}\includegraphics[width=0.77\textwidth]{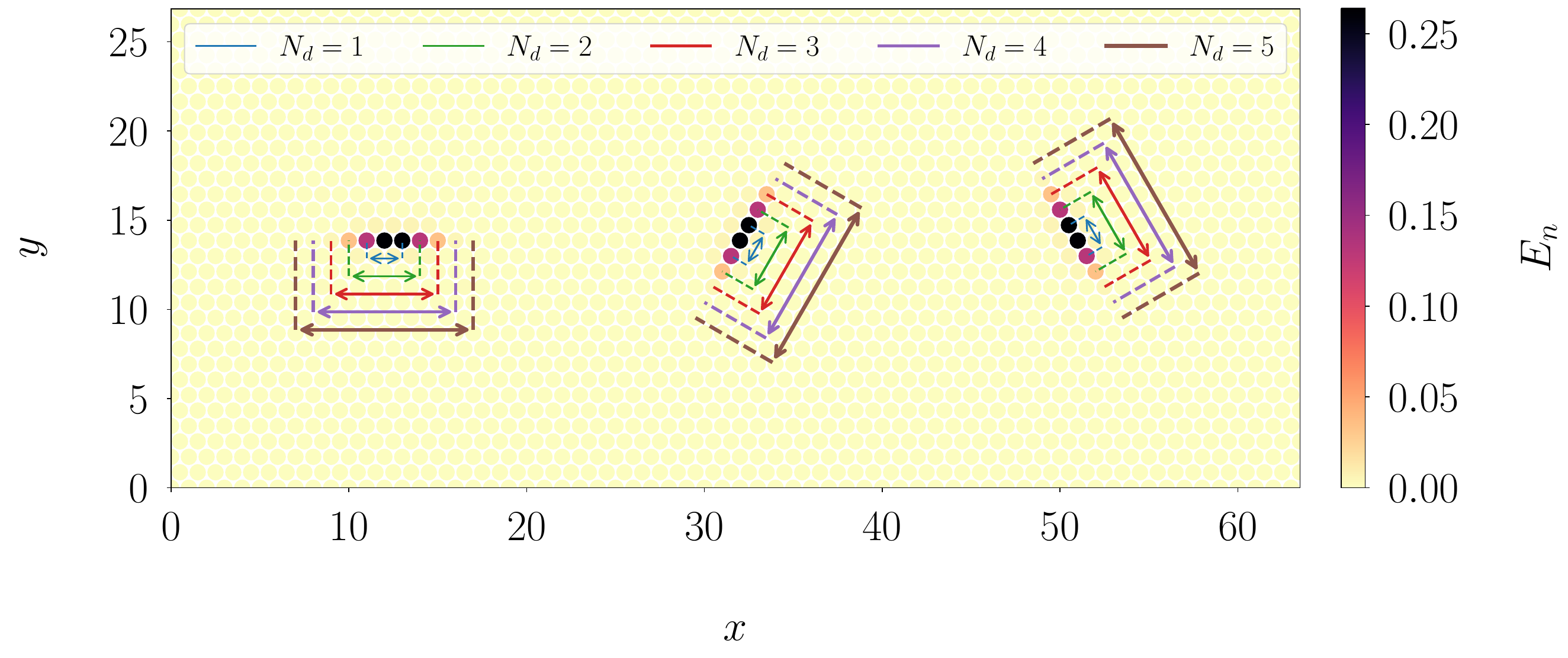}
\caption{1D regions of different directions}
\end{subfigure}

\bigskip

\begin{subfigure}{\textwidth}
\hspace{6em}\includegraphics[width=0.77\textwidth]{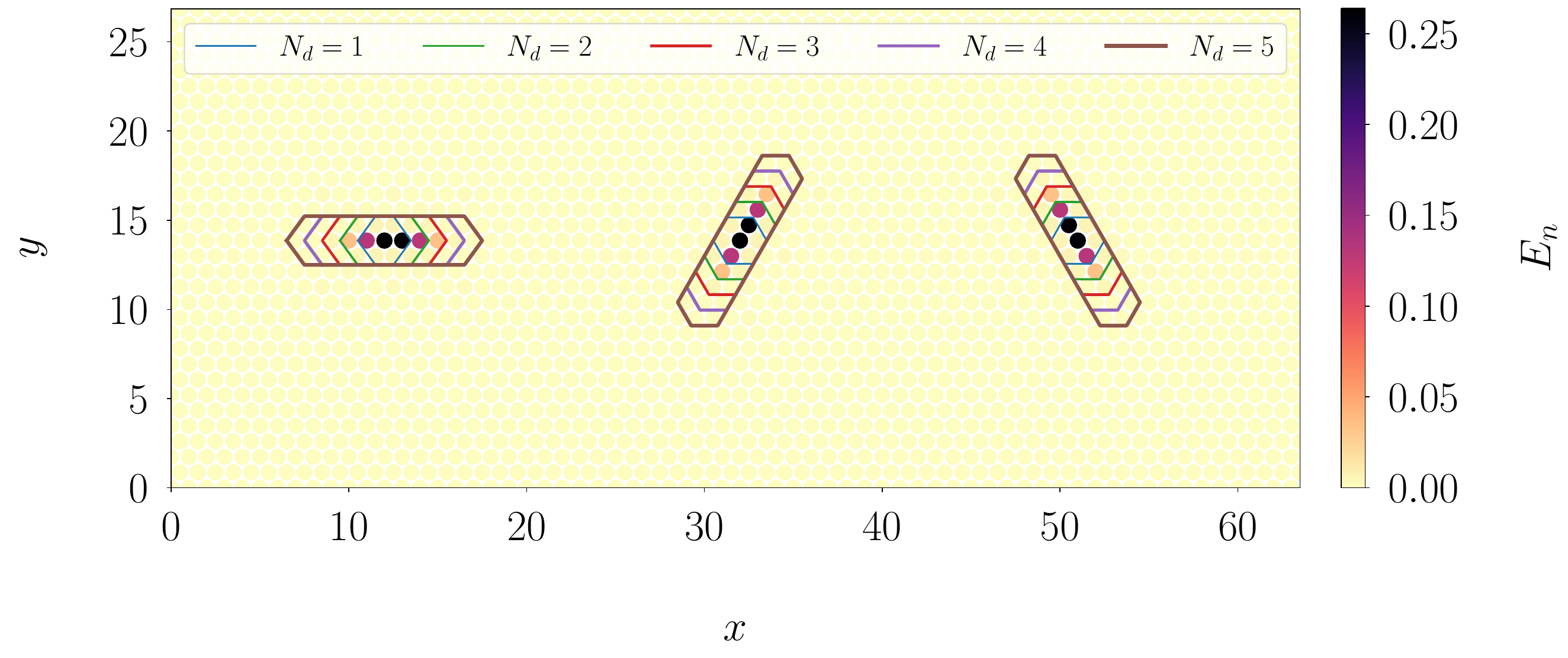}
\caption{Quasi-1D regions of different directions}
\end{subfigure}

\caption{Three different data collection regions: 2D, 1D, and quasi-1D, with the radius parameter $N_d$ and three different directions in the later two cases. The regions with the parameter $N_d$ values ranging from one to five are shown with blue, green, red, purple, and brown lines of increasing thickness, respectively.}\label{fig:data_collection_regions}

\end{figure}

The particle energy density function
\[
E_n=\frac{1}{2}\|\dot{\mathbf{q}}_n\|^2+U(\mathbf{q}_n)+\frac{1}{2}\sum_{l\in\mathcal{L}_n}V(r_{n,l})+\frac{\varepsilon}{2}\geq 0
\]
is defined, which allows for examining the distribution of energy in the lattice in each simulation time step. Three different types of data collection regions are proposed, including two based not only on the structure of the hexagonal lattice but on the geometrical properties of the DBs as well. They are shown in Fig.~\ref{fig:data_collection_regions} for different DB directions, where the parameter $N_d$ is the radius of the corresponding region. A simulation of three stationary DBs with different directions and $\gamma=0.45$ is performed, and the energy density values $E_n$ at the final computational time $T_{end}=10$ are demonstrated, where the number of particles in each lattice row is $N_x=64$, the number of particles in each lattice column is $N_y=32$, and, therefore, the total number of lattice particles is $N=N_x\cdot N_y=2048$. In Fig.~\ref{fig:data_collection_regions}(a), the 2D hexagonal regions are illustrated. Taking into account that the energy of a DB is mostly concentrated only on one layer of particles~\cite{bajars2015nonlinear}, it can be useful to collect data only from this layer, which motivates the consideration of 1D data collection regions, shown in Fig.~\ref{fig:data_collection_regions}(b). However, it is known that DBs are not completely 1D~\cite{bajars2015nonlinear}. Therefore, quasi-1D regions, shown in Fig.~\ref{fig:data_collection_regions}(c), are also proposed.

For each $N_d\in\{1,\,2,\,3,\,4,\,5\}$ and for each region type, a lattice wave dataset is created, obtaining $15$ different datasets. In every case, a total of $N_{sim}=1000$ crystal lattice simulations are performed, where $N_x=32$, $N_y=16$, and $N=512$, saving particle energy density values $E_n$ at the final computational time $T_{end}=50$ of each simulation. In $75\%$ of the simulations, DBs are generated with an equal proportion of waves of each direction. In the remaining $25\%$ of the simulations, phonon waves are generated. For simulating DBs, different values of $\gamma$ are drawn from the uniform distribution $\mathcal{U}(0.25,\,1)$, where the endpoints of the interval are found empirically so that sufficiently long-living localized waves are generated. Therefore, for each value of $N_d$ and for each type of data collection region, a data matrix $\mathbf{X}\in\mathbb{R}^{N_{sim}\times M_d}$ is obtained, where $M_d$ is the number of particles in the data collection region.

As will be seen in Sections~\ref{sec:classification}~and~\ref{sec:numerical_results}, it is sufficient to include only the energy density values in the datasets to significantly reduce the dimensionality of the datasets, train highly accurate lattice wave classifiers and identify localized wave regions, as well as estimate the directionality of the detected DBs. It was found that including also the particle displacements and momenta values in the datasets, as was done in the 1D case in~\cite{bajars2022data}, does not lead to improvement of the results. Moreover, in some cases, this can even lead to a worse reflection of the DB geometrical properties by the identified localization regions.

\section{Dimensionality reduction and classification}\label{sec:classification}

In this section, \emph{Principal Component Analysis} (PCA) is applied to find low-dimensional representations of the datasets described in Section~\ref{sec:data_collection} preserving $95\%$ of their variance~\cite{jolliffe2016principal}. Highly accurate \emph{Support Vector Machine} classifiers (SVCs) are further trained using the low-dimensional data to distinguish between DB and phonon wave data~\cite{geron2019hands}. The performance of the obtained classifiers is evaluated and compared.

\subsection{Dimensionality reduction with PCA}

In order to enhance the computational efficiency, before training lattice wave classifiers, the dimensionality of the lattice wave datasets is reduced using PCA. The main objective of this method is to extract the so-called \emph{principal components} from the data that preserve a specified amount of variance or information contained in the original high-dimensional dataset. 

To achieve the goal of PCA, the eigen-decomposition of the covariance matrix
\[
\mathbf{C}=\frac{1}{N_{sim}-1}\mathbf{Y}^T\mathbf{Y}
\]
is performed, where $\mathbf{Y}=\mathbf{X}-\bar{\mathbf{X}}$ is the centered data matrix,
\[
\bar{\mathbf{X}}=
\begin{pmatrix}
1 \\
\vdots \\
1
\end{pmatrix}
\begin{pmatrix}
\bar{x}_1,\ldots,\,\bar{x}_{M_d}
\end{pmatrix}\quad\text{and}\quad
\bar{x}_j=\frac{1}{N_{sim}}\sum_{i=1}^{N_{sim}}X_{ij}.
\]
Here, $\mathbf{X}\in\mathbb{R}^{N_{sim}\times M_d}$ is one of the lattice wave data matrices described in Section~\ref{sec:data_collection}. The matrix $\mathbf{C}$ is symmetric and semi-positive definite, which means that it has $M_d$ real non-negative eigenvalues and its eigenvectors are orthogonal. To obtain the low-dimensional representation $\mathbf{Z}\in\mathbb{R}^{N_{sim}\times d}$, the centered data matrix $\mathbf{Y}$ is projected onto the eigenvectors associated with the largest $d\ll M_d$ eigenvalues. The columns of the matrix $\mathbf{Z}$ are the searched principal components, whereas the $d$ largest eigenvalues sorted in descending order correspond to the amount of the original dataset variance preserved by each principal component. It is significant to note that the eigenvalues of the covariance matrix $\mathbf{C}$ are the squared singular values of the matrix $\mathbf{Y}/\sqrt{N_{sim}-1}$, which highlights a strong connection between PCA and Singular Value Decomposition~\cite{jolliffe2016principal}.

The number of dimensions needed to preserve $95\%$ of the original dataset variance is shown in Table~\ref{tab:PCA}. The dimension numbers of the low-dimensional datasets are shown together with the dimension numbers of the original datasets and the approximate percentage by which the dimension number can be reduced, preserving $95\%$ of the variance (in brackets). In order to preserve $95\%$ of the original dataset variance, it is necessary to include in the low-dimensional dataset at most three principal components. Hence, it is possible to reduce the dimension number of the original dataset by up to approximately $97\%$, which is a significant improvement.

\begin{table}[H]
\caption{Dimension numbers needed to preserve $95\%$ of the original dataset variance applying PCA.}
\label{tab:PCA}
\begin{center}
\begin{tabular}{lllllll}
\hline
   & $N_d=1$ & $N_d=2$ & $N_d=3$ & $N_d=4$ & $N_d=5$ \\ \hline
2D & $3$ / $7$ ($57\%$) & $3$ / $19$ ($84\%$) & $3$ / $37$ ($92\%$) & $3$ / $61$ ($95\%$) & $3$ / $91$ ($97\%$) \\
1D & $1$ / $3$ ($67\%$) & $1$ / $5$ ($80\%$) & $1$ / $7$ ($86\%$) & $1$ / $9$ ($89\%$) & $1$ / $11$ ($91\%$)  \\
Quasi-1D & $1$ / $7$ ($86\%$) & $1$ / $13$ ($92\%$) & $1$ / $19$ ($95\%$) & $1$ / $25$ ($96\%$) & $1$ / $31$ ($97\%$)  \\ \hline
\end{tabular}
\end{center}
\end{table}

\subsection{SVM classification}

After reducing the dimensionality of the $15$ datasets described in Section~\ref{sec:data_collection} with PCA to the minimal number of dimensions needed to preserve $95\%$ of the original dataset variance, for each of the low-dimensional datasets, a nonlinear SVC is trained to classify lattice waves using locally sampled data. The objective is to obtain highly accurate binary classifiers that can determine whether a specific data instance corresponds to a DB.

The process of training the SVCs is described in Appendix~\ref{ap:SVC_training}. For each of the low-dimensional datasets, a nonlinear SVC is trained using the Gaussian \emph{Radial Basis Function} (RBF) kernel:
\[
\mathcal{K}(\mathbf{z}^i,\,\mathbf{z}^j)=\exp(-\zeta\|\mathbf{z}^i-\mathbf{z}^j\|^2),
\]
where $\zeta>0$ and $\mathbf{z}^i,\,\mathbf{z}^j\in\mathbb{R}^d$ are respectively the $i$-th and $j$-th data instances of a low-dimensional dataset $\mathbf{Z}\in\mathbb{R}^{N_{sim}\times d
}$, $i,\,j=1,\ldots,\,N_{sim}$~\cite{geron2019hands}. For evaluating the performance of the obtained classifiers, such binary classification metrics as precision ($\nu_P$) and recall ($\nu_R$) are used:
\[
\nu_P=\frac{n_{TP}}{n_{FP}+n_{TP}}\quad\text{and}\quad\nu_R=\frac{n_{TP}}{n_{FN}+n_{TP}},
\]
where $n_{TP}$ is the number of true positives (positives classified as positives), $n_{FP}$ is the number of false positives (negatives classified as positives), $n_{FN}$ is the number of false negatives (positives classified as negatives), assuming that the DB data belongs to the positive class and the phonon wave data belongs to the negative class. For training the classifiers, $70\%$ of the data are used, and, for testing them, the remaining $30\%$ of the data are used. Values of the SVC regularization hyperparameter $\lambda$ (see Appendix~\ref{ap:SVC_training}), found with the grid search technique, are shown in Table~\ref{tab:SVC_lambdas}, and a problem scaled value of the parameter $\zeta$ is used: $\zeta=1/(d\cdot v)$, where $d$ is the dataset dimension number and $v$ is its variance~\cite{geron2019hands}.

\begin{table}[H]
\caption{Values of the SVC regularization hyperparameter $\lambda$ which are found using the grid search technique.}
\label{tab:SVC_lambdas}
\begin{center}
\begin{tabular}{lllllll}
\hline
   & $N_d=1$ & $N_d=2$ & $N_d=3$ & $N_d=4$ & $N_d=5$ \\ \hline
2D & $1$ & $1$ & $1$ & $1$ & $1$ \\
1D & $0.01$ & $0.01$ & $0.01$ & $0.01$ & $0.01$ \\
Quasi-1D & $0.01$ & $0.01$ & $0.01$ & $0.01$ & $0.01$ \\ \hline
\end{tabular}
\end{center}
\end{table}

To analyze and compare the performance of the trained classifiers, $100$ random train-test splits of each dataset are performed. Precision and recall are evaluated on the test sets and on the validation sets, which are obtained by applying $K$-fold cross-validation with $K=5$ to the training data~\cite{geron2019hands}. The average precision and recall values are shown in Table~\ref{tab:SVC_results}. The results are excellent in all cases, which means that highly accurate lattice wave data classifiers are obtained. In Section~\ref{sec:numerical_results}, the ability of the classifiers to find localized wave regions in crystal lattice numerical simulations is analyzed.

\begin{table}[H]
\caption{Average precision and recall values obtained by training the lattice wave classifiers on different datasets. In each case, $100$ random train-test splits are performed, and cross-validation is applied to the training data.}
\label{tab:SVC_results}
\begin{center}
\begin{tabular}{lllllll}
\hline
               &  & $N_d=1$ & $N_d=2$ & $N_d=3$ & $N_d=4$ & $N_d=5$ \\ \hline
\multicolumn{1}{l}{\multirow{4}{*}{2D}}   & Precision (validation set) & $1$ & $1$ & $1$ & $1$ & $1$ \\
\multicolumn{1}{l}{}                      & Recall (validation set) & $1$ & $1$ & $1$ & $1$ & $1$ \\
\multicolumn{1}{l}{}                      & Precision (testing set) & $1$ & $1$ & $1$ & $1$ & $1$ \\
\multicolumn{1}{l}{}                      & Recall (testing set) & $1$ & $1$ & $1$ & $1$ & $1$ \\ \cmidrule{1-7}
\multicolumn{1}{l}{\multirow{4}{*}{1D}}   & Precision (validation set) & $1$ & $1$ & $1$ & $1$ & $1$ \\
                                          & Recall (validation set) & $1$ & $1$ & $1$ & $1$ & $1$ \\
                                          & Precision (testing set) & $1$ & $1$ & $1$ & $1$ & $1$ \\
                                          & Recall (testing set) & $0.9926$ & $0.9925$ & $0.9926$ & $0.9925$ & $0.9926$ \\ \cmidrule{1-7}
\multicolumn{1}{l}{\multirow{4}{*}{Quasi-1D}} 
										   & Precision (validation set) & $1$ & $1$ & $1$ & $1$ & $1$ \\ 
                                          & Precision (testing set) & $1$ & $1$ & $1$ & $1$ & $1$ \\
                                          & Recall (validation set) & $1$ & $1$ & $1$ & $1$ & $1$ \\
                                          & Recall (testing set) & $0.9926$ & $0.9926$ & $0.9927$ & $0.9927$ & $0.9927$ \\ \hline
\end{tabular}
\end{center}
\end{table}

\FloatBarrier

\section{Numerical results}\label{sec:numerical_results}

In this section, the lattice wave data classifiers, described in Section~\ref{sec:classification}, are applied in crystal lattice numerical simulations to find localized wave regions. Moreover, an algorithm that allows for estimating DB directionality is proposed and tested.

\subsection{Localization region detection}

To find localization regions in a crystal lattice simulation, the sliding window approach is applied, as was done in~\cite{bajars2022data}. A window moves along the lattice and makes predictions using one of the previously trained classifiers. If a localized wave is detected for a specific position of the window, then it is considered that all particles inside the window belong to the localization region. After the window passes through the entire lattice, each particle is either detected to belong to a localization region or detected not to belong to any localization region. Thus, the localization regions are found. In contrast to the 1D case, sliding windows of different shapes can be used. In the 2D case, the sliding window shape coincides with the shape of the data collection region, which was used for training the applied classifier, see Fig.~\ref{fig:data_collection_regions}.

\begin{figure}[t]

\centering

\begin{subfigure}{0.7\textwidth}
\centering
\includegraphics[width=\textwidth]{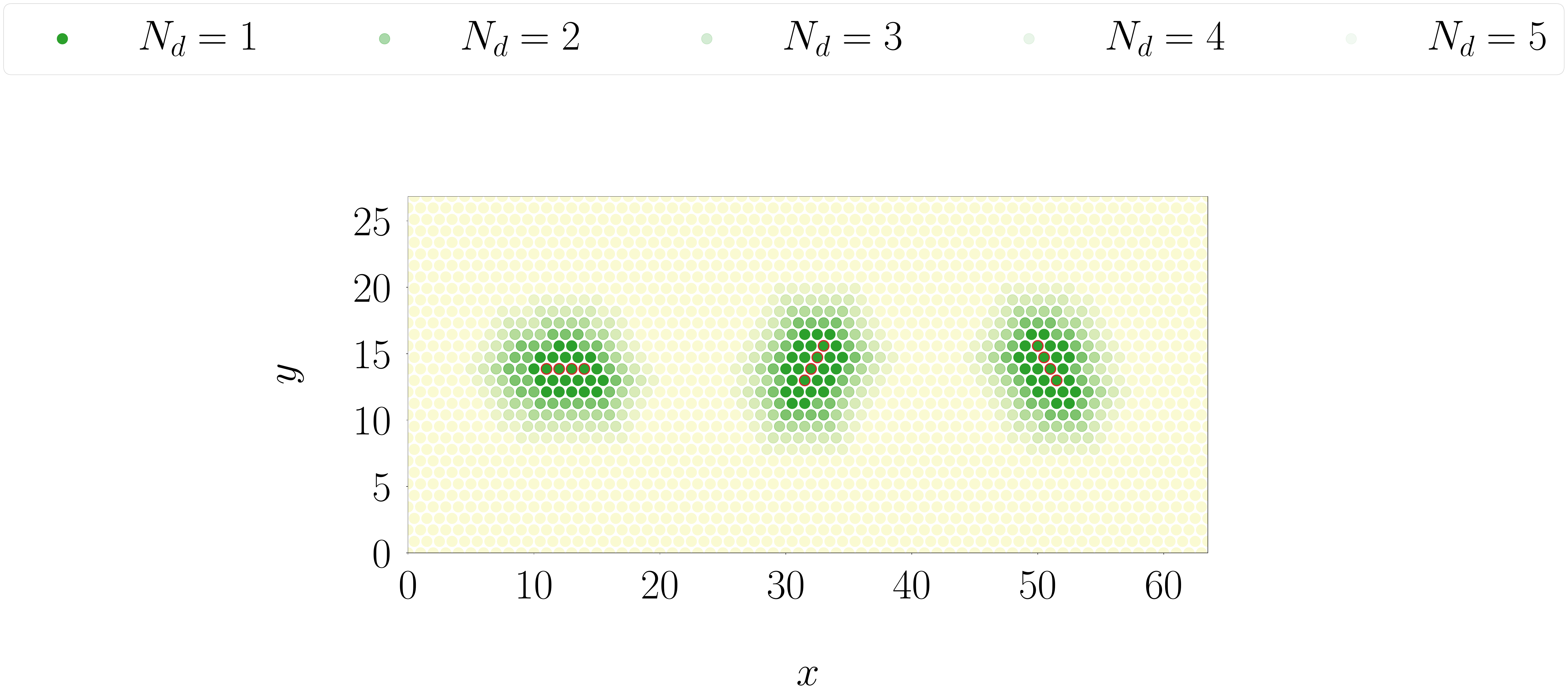}
\caption{2D window}
\end{subfigure}
        
\bigskip
        
\begin{subfigure}{0.45\textwidth}
\centering 
\includegraphics[width=\textwidth]{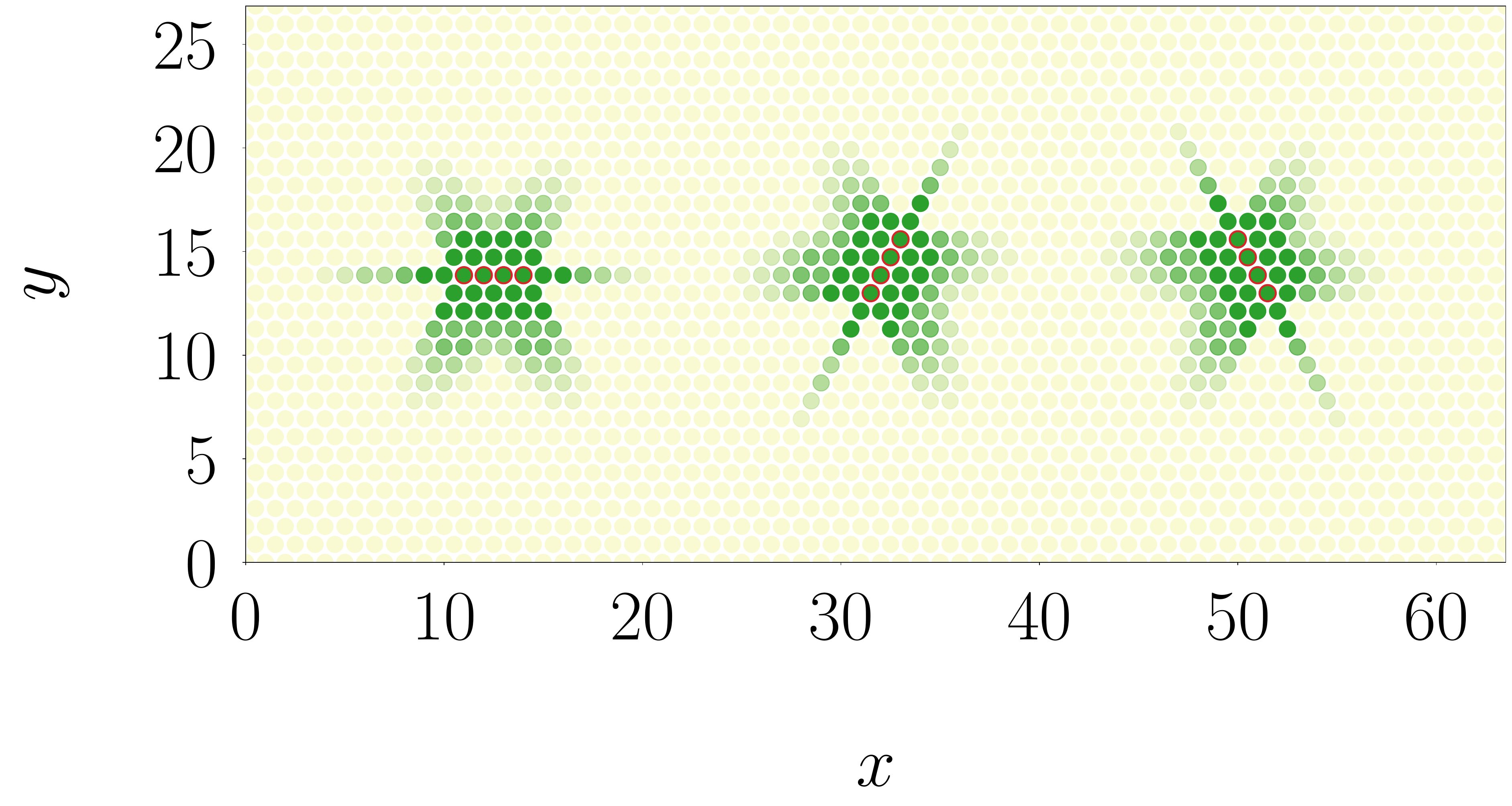}
\caption{1D windows (summation)}
\end{subfigure}
\hspace{4em}
\begin{subfigure}{0.45\textwidth}
\centering 
\includegraphics[width=\textwidth]{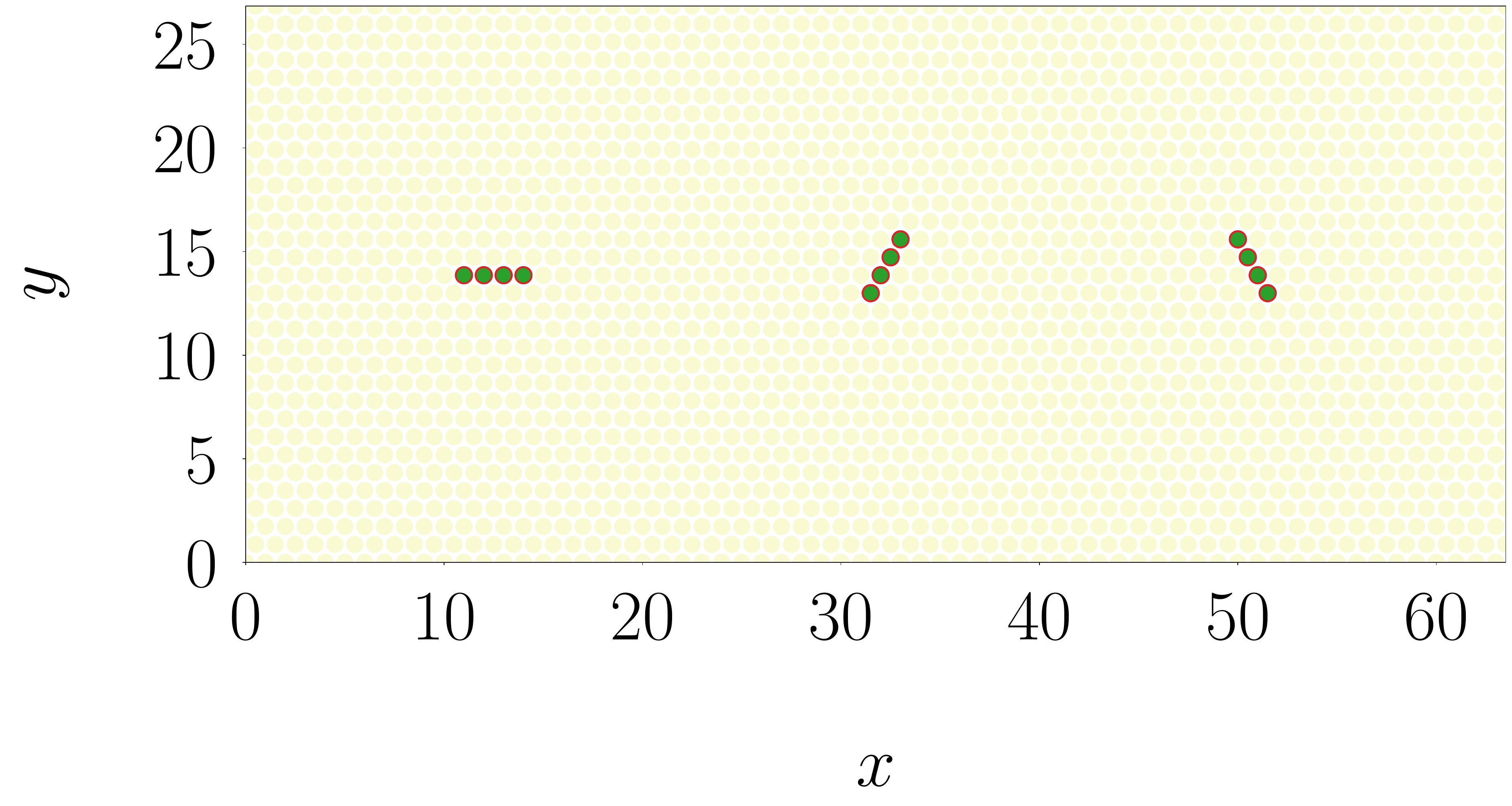}
\caption{1D windows (product)}
\end{subfigure}

\bigskip        
        
\begin{subfigure}{0.45\textwidth}
\centering 
\includegraphics[width=\textwidth]{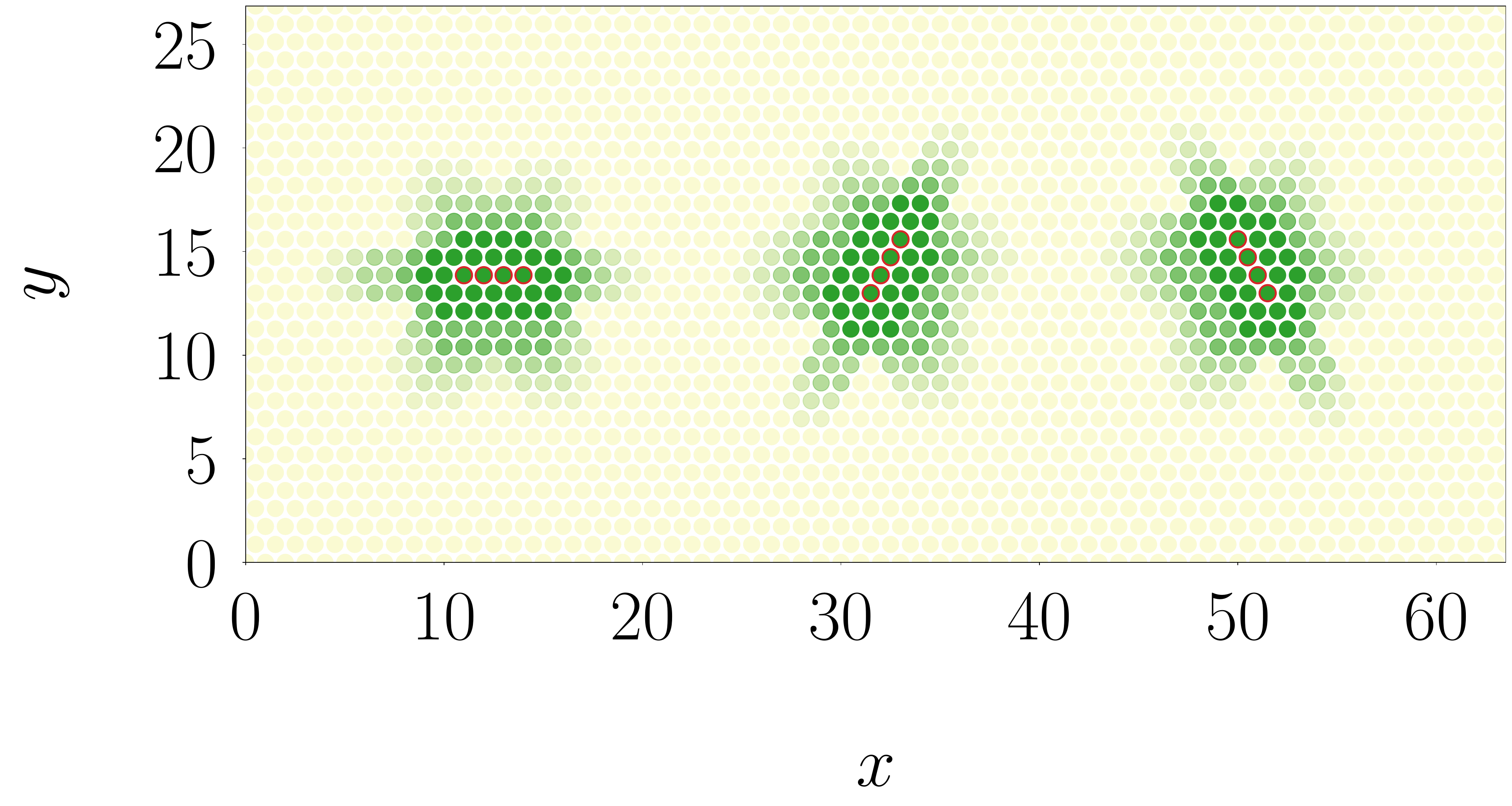}
\caption{Quasi-1D windows (summation)}
\end{subfigure}
\hspace{4em}
\begin{subfigure}{0.45\textwidth}
\centering 
\includegraphics[width=\textwidth]{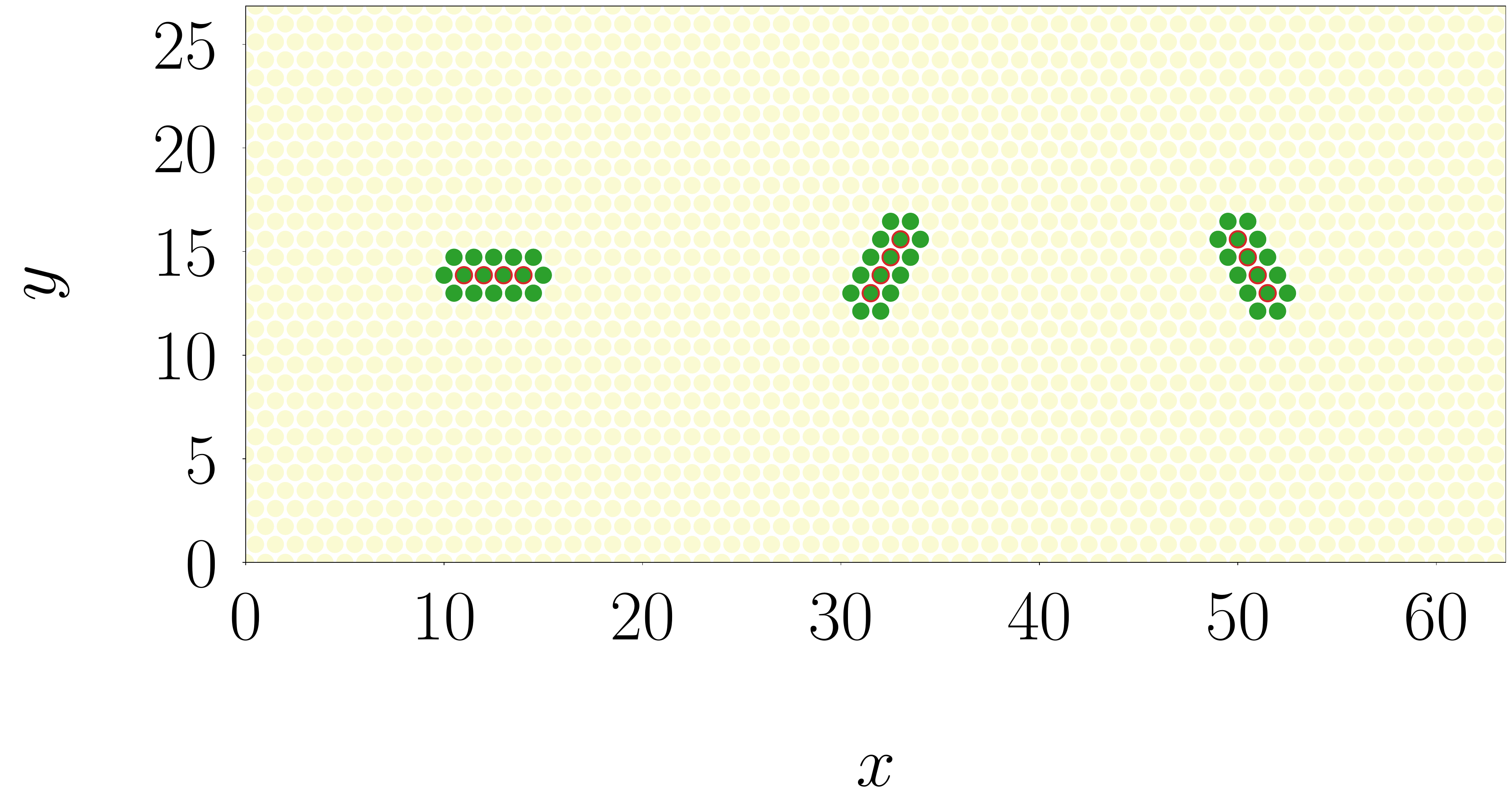}
\caption{Quasi-1D windows (product)}
\end{subfigure}

\caption{Localization regions detected using the 2D, 1D, and quasi-1D sliding windows of different sizes and three different directions in the latter two cases, aggregating the predictions with the summation and product approaches. The localization regions obtained with the window radius parameter $N_d$ values ranging from one to five are shown with a green color scale of decreasing intensity as the $N_d$ value increases.}\label{fig:loc_regions}

\end{figure}

Trying to identify a localized wave region, the directionality is usually unknown. Therefore, in the 2D case, when the data is collected from the 1D or quasi-1D regions, instead of one window, three windows of different directions move along the lattice simultaneously, making predictions. As a result, for each lattice particle, three predictions are obtained, corresponding to different crystallographic directions~\eqref{eq:directions}, which have to be aggregated. Two different approaches for aggregating the predictions are considered (see Fig.~\ref{fig:loc_regions}). In the first approach, which is called the \emph{summation approach}, a particle belongs to a localization region if at least one of the windows indicates so. In contrast, when applying the \emph{product approach}, a particle is considered to belong to a localization region if it belongs to this region according to each of the windows that identified it. Thus, if, for instance, only one of the windows was unable to identify a localization region, then its prediction is not taken into account, and the predictions of the remaining two windows are aggregated.

In Fig.~\ref{fig:loc_regions}, the localization regions are shown, which are obtained using the energy density data collected from the regions shown in Fig.~\ref{fig:data_collection_regions}, simulating three stationary DBs with the same parameters as in Section~\ref{sec:data_collection}. The red circles denote the lattice region where the corresponding DB is initialized. In the case of the 1D and quasi-1D regions, two aggregation approaches are applied to the predictions obtained with sliding windows of different directions: in Fig.~\ref{fig:loc_regions}(b) and Fig.~\ref{fig:loc_regions}(d), the results obtained with the summation approach are shown, while, in Fig.~\ref{fig:loc_regions}(c) and Fig.~\ref{fig:loc_regions}(e), the results obtained with the product approach are shown. It can be seen that, independently of the data collection region considered, in all the cases, the regions of localization are found. However, not always are the geometric properties of DBs described in Section~\ref{sec:mathematical_model} preserved by the identified localization regions. It is known that DBs are quasi-1D, which is not reflected in the case of the 2D and 1D windows, as well as in the case of the quasi-1D windows applied together with the summation approach. In contrast, using the data from the quasi-1D regions together with the product approach allows for obtaining localization regions that are quasi-1D. In this case, the directionality of the DBs can be determined empirically. It can also be noted that, in the cases when the product approach is applicable and used, the maximal size of the detected localization regions is smaller than the maximal size of the localization regions detected applying the 2D sliding windows, as well as the 1D and quasi-1D sliding windows with the summation approach. Moreover, despite the localization regions obtained with the 1D windows and the product approach not being quasi-1D, the energy distribution in the lattice is captured by the regions more precisely in this case than in the other cases, as can be seen from the energy density values of the lattice particles shown in Fig.~\ref{fig:data_collection_regions}.

\subsection{Localization region segmentation and directionality estimation}

After identifying localized wave regions in a crystal lattice simulation at a given time, it is also useful to detect the crystallographic directions of the DBs corresponding to the found regions of localization. It was shown that the regions themselves can reveal the geometric properties of the corresponding waves, and sometimes the directionality can be determined empirically. However, performing a simulation with a large number of time steps, it is inefficient to analyze crystal lattice classification data empirically. Therefore, one of the goals of this work is to present a highly accurate algorithm to identify a DB crystallographic direction based on the found localization region.

In order to analyze individual localization regions, the classification data has to be segmented. The objective is to combine particles that belong to the same region of localization into one group, forming groups of particles corresponding to different localization regions. Then, every single region can be studied. It is assumed that, in a fixed time step of a crystal lattice simulation, localization regions are detected but not segmented. Thus, the set $\mathcal{G}$ is obtained, containing indices of the particles that belong to the regions of localization. In this work, the recursive Algorithm~\ref{alg:segmentation} is considered for localization region segmentation. If the localization regions are detected, then, after applying the algorithm, the sets $\mathcal{R}_1,\ldots,\,\mathcal{R}_{N_r}$ are obtained, where $\mathcal{R}_i$ contains indices of the particles belonging to the $i$-th localization region and $N_r$ is the number of detected localization regions, $i=1,\ldots,\,N_r$.

\begin{algorithm}[t]
\caption{Localization region segmentation}\label{alg:segmentation}
\begin{enumerate}
\item If $\mathcal{G}=\emptyset$, then no localization regions are detected and the algorithm is terminated. Otherwise, Steps 2--6 are sequentially executed.
\item Let $i=1$.
\item Let $\mathcal{R}_i=\emptyset$.
\item An index $n_i\in\mathcal{G}$ is added to the set $\mathcal{R}_i$ and removed from the set $\mathcal{G}$.
\item The set $\mathcal{R}_i$ is recursively augmented with those particle indices from $\mathcal{G}$ which are direct neighbors of the particles in $\mathcal{R}_i$ removing the indices from the set $\mathcal{G}$. The process is terminated when there are no neighbors of particles in $\mathcal{R}_i$ that belong to $\mathcal{G}$.
\item If $\mathcal{G}=\emptyset$, then localization regions are segmented and the algorithm is terminated. Otherwise, $i$ is incremented by one and Steps 3--6 are repeated.
\end{enumerate}
\end{algorithm}

The next objective is to determine the directionality of the localized wave corresponding to a detected region of localization if the indices of the particles belonging to this region are known. The method proposed in this work is based on the knowledge of the considered region's geometric properties and the energy distribution in the lattice. Firstly, the general notion of a coefficient of correspondence of a set to a direction in the plane based on the energy distribution is introduced, which is then applied to the case of the crystal lattice model. A non-empty, compact, and connected set $\mathcal{S}\subset\mathbb{R}^2$ is considered together with a continuous normalized energy density function $\bar{E}\colon\mathcal{S}\to[0,1]$. The angle $\alpha\in[0,\pi)$ defines a direction in the plane. It is assumed that $\iint_{\mathcal{S}}\bar{E}(x,y)\,dxdy>0$. The \emph{coefficient of correspondence of $\mathcal{S}$ to the direction $\alpha$} is defined as follows:
\[
I_{\alpha}=\frac{\iint_{\mathcal{S}} \bar{E}(x,y)\exp{(-\kappa d_{\alpha}(x,y))}\,dxdy}{\iint_{\mathcal{S}}\bar{E}(x,y)\,dxdy}\in(0,1],
\]
where $\kappa>0$ and $d_{\alpha}(x,y)$ is the Euclidean distance from a point $(x,y)$ to the line with angle $\alpha$ to which the geometric center of $\mathcal{S}$ belongs.

Applying the previously introduced notion of a set correspondence to a direction in the plane in the case of the crystal lattice particle data, the direction of each localization region detected in a simulation can be found. It is assumed that $\mathcal{R}$ is a set obtained with Algorithm~\ref{alg:segmentation} containing the particle indices that belong to one of the detected localization regions. The coefficients
\begin{equation}\label{eq:correspondence_coefficients}
\hat{I}_{\alpha}=\frac{\sum_{n\in\mathcal{R}} \bar{E}_n\exp{(-\kappa d_{\alpha,n}})}{\sum_{n\in\mathcal{R}}\bar{E}_n},\quad\text{for}\quad\alpha\in\left\{0,\,\pi/3,\,2\pi/3\right\},
\end{equation}
are computed, where $\bar{E}_n=E_n/\max_{n\in\mathcal{R}} E_n$, $d_{\alpha,n}=d_{\alpha}(q_{0nx},\,q_{0ny})$ and $\mathbf{q}_{0n}=(q_{0nx},\,q_{0ny})^T\in\mathbb{R}^2$ is the equilibrium position of the particle with index $n$. The direction of the localized wave region is defined by the angle $\alpha\in\left\{0,\,\pi/3,\,2\pi/3\right\}$ corresponding to the highest value of $\hat{I}_{\alpha}$. Therefore, to determine the directionality of the DB corresponding to the detected localization region, it is necessary to compute $\hat{I}_0$, $\hat{I}_{\pi/3}$ and $\hat{I}_{2\pi/3}$ and choose the direction associated with the maximal coefficient. In this work, if the difference between the maximal coefficient and the second biggest coefficient is smaller than $0.001$, then it is assumed that the considered algorithm is unable to determine the localized wave directionality.

For performance evaluation of the proposed directionality detection algorithm, numerical simulations of four different DBs are run: a simulation of a weak stationary DB, a simulation of a strong stationary DB, a simulation of a weak traveling DB, and a simulation of a strong traveling DB. The weak localized waves are generated using $\gamma=0.36$, while the strong localized waves are generated using $\gamma=-0.5$. The stationary DBs have the direction corresponding to $\alpha=\pi/3$, and the traveling DBs have the direction corresponding to $\alpha=0$. All the simulations are performed till $T_{end}=100$, i.e., running $10000$ time steps, and considering the crystal lattice model with $N_x=64$ and $N_y=32$. In each simulation time step, the localization region is found using the classifiers trained in Section~\ref{sec:classification} together with the sliding window approach, and then the directionality of the corresponding DB is detected. If the localization region is not found, then it is assumed that its directionality is not identified. After performing the simulations, the directionality detection accuracy is computed by dividing the number of times when the directionality is identified correctly by the total number of simulation time steps. For finding the region of localization, sliding windows of different shapes and sizes are used together with different prediction aggregation techniques when required.

\begin{figure}[t]

\centering

\begin{subfigure}{0.9\textwidth}
\centering
\includegraphics[width=\textwidth]{Figure_3_legend.pdf}
\end{subfigure}
        
\bigskip

\begin{subfigure}{0.4\textwidth}  
\centering 
\includegraphics[width=\textwidth]{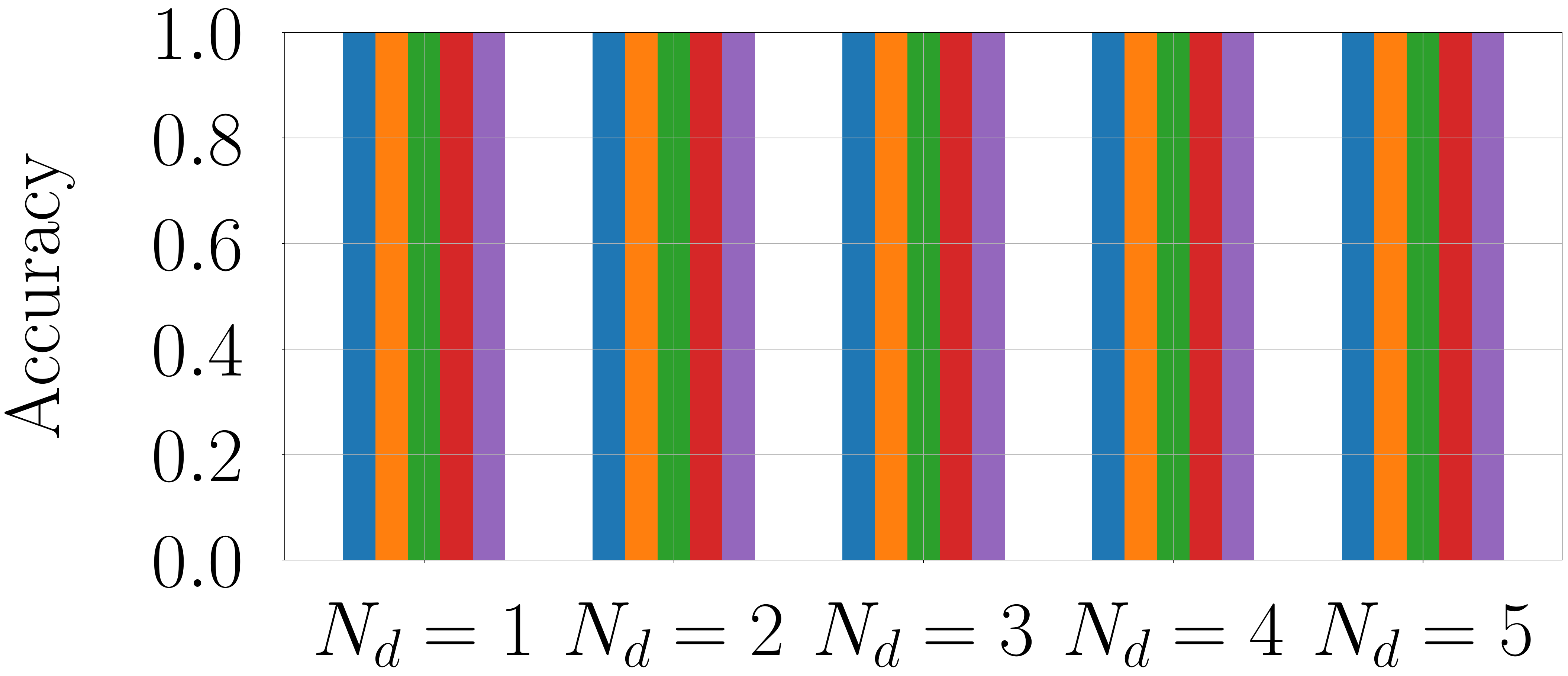}
\caption{Stationary DB with $\gamma=0.36$ and $\alpha=\pi/3$}
\end{subfigure}
\hspace{4em}
\begin{subfigure}{0.4\textwidth}   
\centering 
\includegraphics[width=\textwidth]{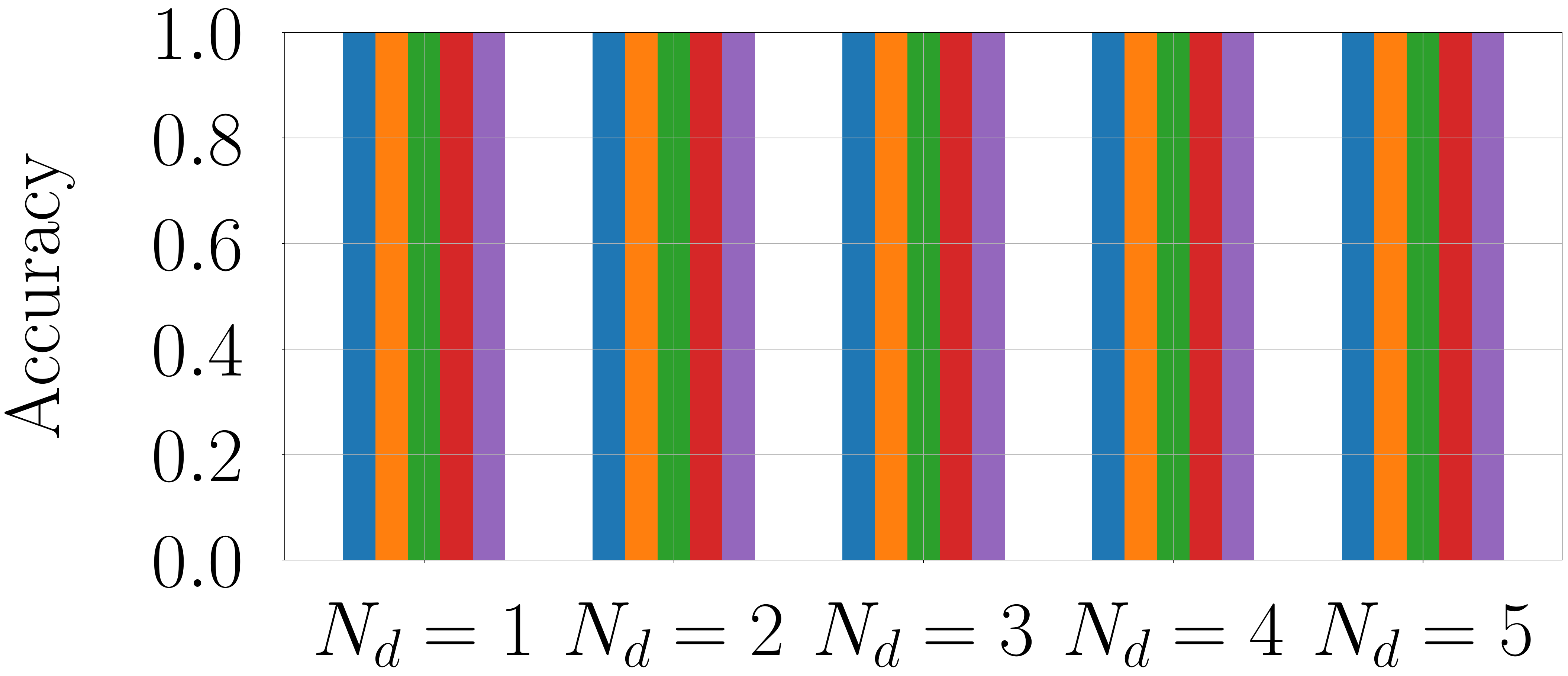}
\caption{Stationary DB with $\gamma=-0.5$ and $\alpha=\pi/3$}
\end{subfigure}

\bigskip        
        
\begin{subfigure}{0.4\textwidth}  
\centering 
\includegraphics[width=\textwidth]{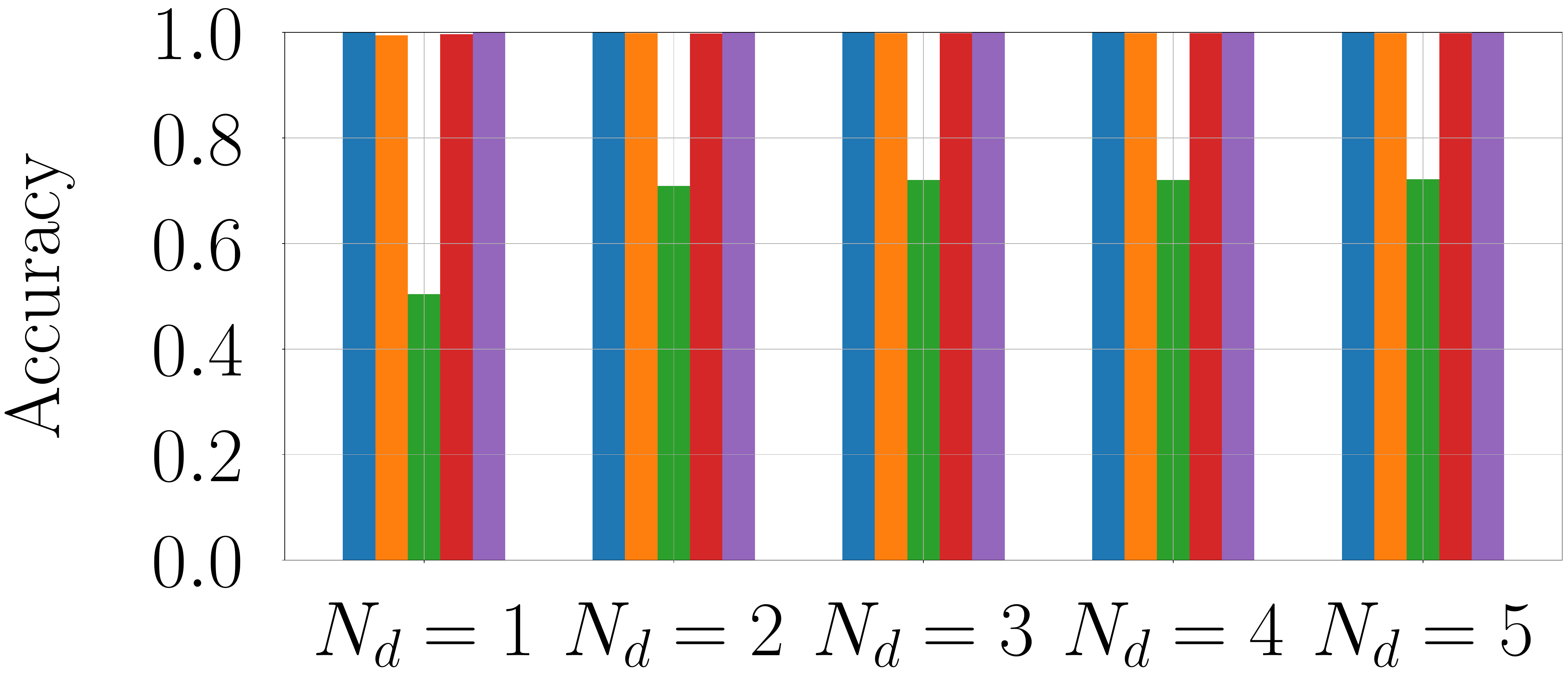}
\caption{Traveling DB with $\gamma=0.36$ and $\alpha=0$}
\end{subfigure}
\hspace{4em}
\begin{subfigure}{0.4\textwidth}   
\centering 
\includegraphics[width=\textwidth]{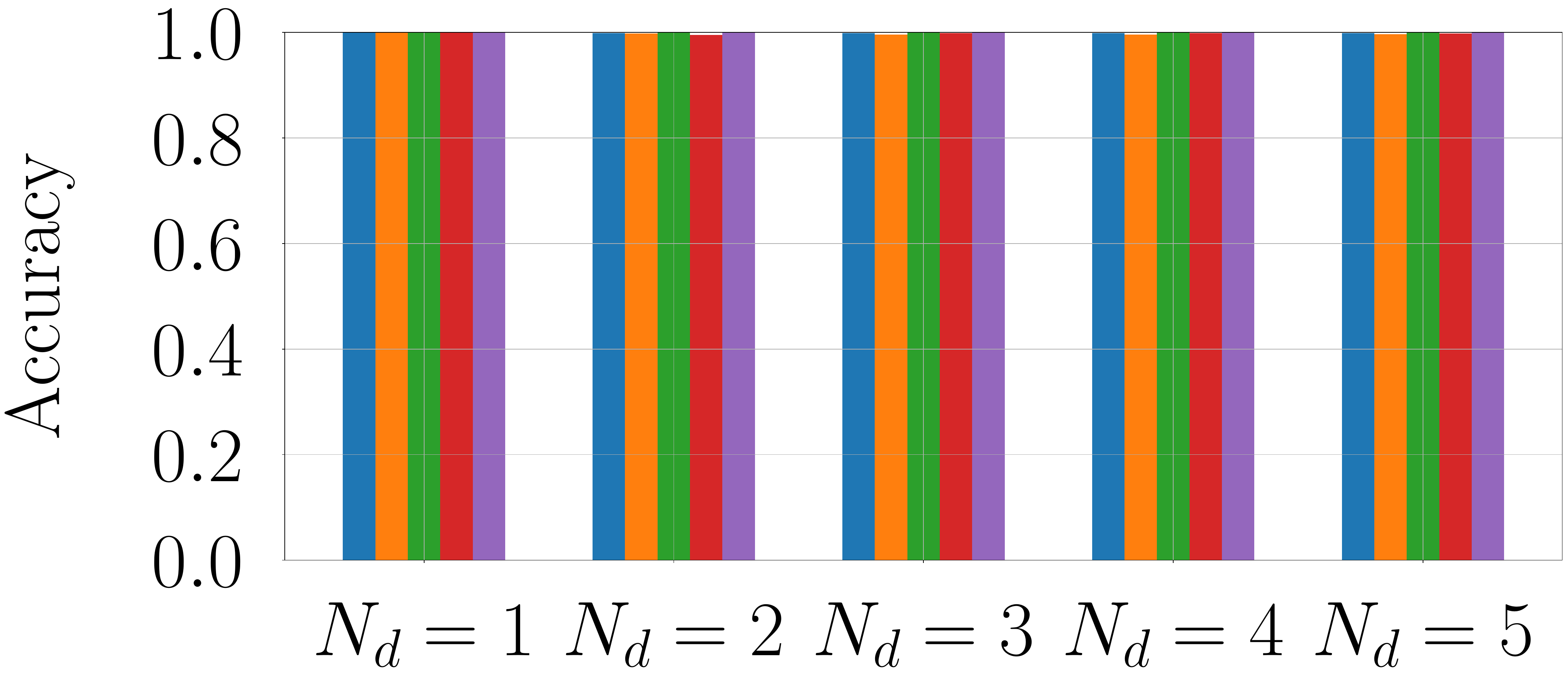}
\caption{Traveling DB with $\gamma=-0.5$ and $\alpha=0$}
\end{subfigure}

\caption{Accuracy of localization region directionality estimation with $\kappa=0.1$, detecting the regions of localization using the 2D, 1D, and quasi-1D sliding windows of different sizes and three different directions in the latter two cases, aggregating the predictions with the summation and product approaches. The results corresponding to the 2D sliding windows, 1D sliding windows with the summation and product approaches, and quasi-1D sliding windows with the summation and product approaches are shown in blue, orange, green, red, and purple, respectively.}\label{fig:direction_detection_accuracy}

\end{figure}

To find the optimal value of the parameter $\kappa$ for computing the coefficients of correspondence~\eqref{eq:correspondence_coefficients}, the grid search technique is applied, performing the simulations with $\kappa\in\{0.01,\,0.1,\,1,\,10,\,100\}$ and choosing the value corresponding to the highest directionality detection accuracy~\cite{geron2019hands}. It was found that the highest accuracy values, which are shown in Fig.~\ref{fig:direction_detection_accuracy}, are obtained with $\kappa=0.1$.

As can be seen in Fig.~\ref{fig:direction_detection_accuracy}, for all sizes and shapes of the sliding windows as well as for both prediction aggregation approaches when they are used, the directionality of the stationary DBs is correctly identified in all simulation time steps. The directionality of the strong traveling DBs is also identified with high accuracy, which exceeds $0.99$ in all the considered cases. In contrast, the accuracy of the directionality detection of the weak traveling DB differs using different sliding windows. Applying the 2D window, 1D windows with the summation approach, and quasi-1D windows, the accuracy is still higher than $0.99$, while applying the 1D window with the product approach, the accuracy does not exceed $0.73$. 

It can be concluded that it is possible to detect localized wave region directionality with high accuracy using the proposed algorithm, and all shapes as well as all considered sizes of the sliding windows, except the 1D window with the product approach, are suitable for this purpose. It is also worth noting that, in all the considered cases, the localization regions are correctly found in all simulation time steps.

\subsection{Numerical study of discrete breather interactions}

In this section, the algorithms for detecting localization regions and estimating the directionality of the corresponding localized waves, which were proposed earlier, are applied to crystal lattice simulations where many DBs propagate and interact. The main objective of such a simulation study is to apply methods tested in ideal conditions in a more complex setting, which allows for evaluation of their generalization ability and, therefore, provides a more reliable view of their performance. Moreover, the numerical study of two traveling DB head-to-head and head-to-adjacent-head collisions is performed, providing insights about the localization region transformations during these collisions as well as the energy exchange between the waves, which leads to the formation of DBs with different properties.

As was found earlier in the current section, in contrast to other data collection regions, collecting data from the quasi-1D regions and using the quasi-1D sliding windows together with the product approach to later find regions of localization in numerical simulations, allows for preserving such an important geometric property of DBs as quasi-one dimensionality. Moreover, the localization regions found in this way are smaller, which provides an opportunity to locate the regions of localization more precisely. It was also shown that using such a shape of the sliding windows allows for detecting the underlying localized wave directionality with high accuracy. On the other hand, using only one sliding window instead of three, which can be achieved by applying the 2D sliding window, is a computationally cheaper task. Furthermore, this sliding window shape also allows for identifying localized wave directionality with high precision. Taking this into account, firstly, two of the approaches mentioned above are applied to a DB interaction simulation and compared, evaluating their performance and examining whether the advantages of the quasi-1D shape are essential in practical applications.

\begin{figure}[t]
\centering
\includegraphics[trim={0 0 0 5cm}, clip, width=0.77\linewidth]{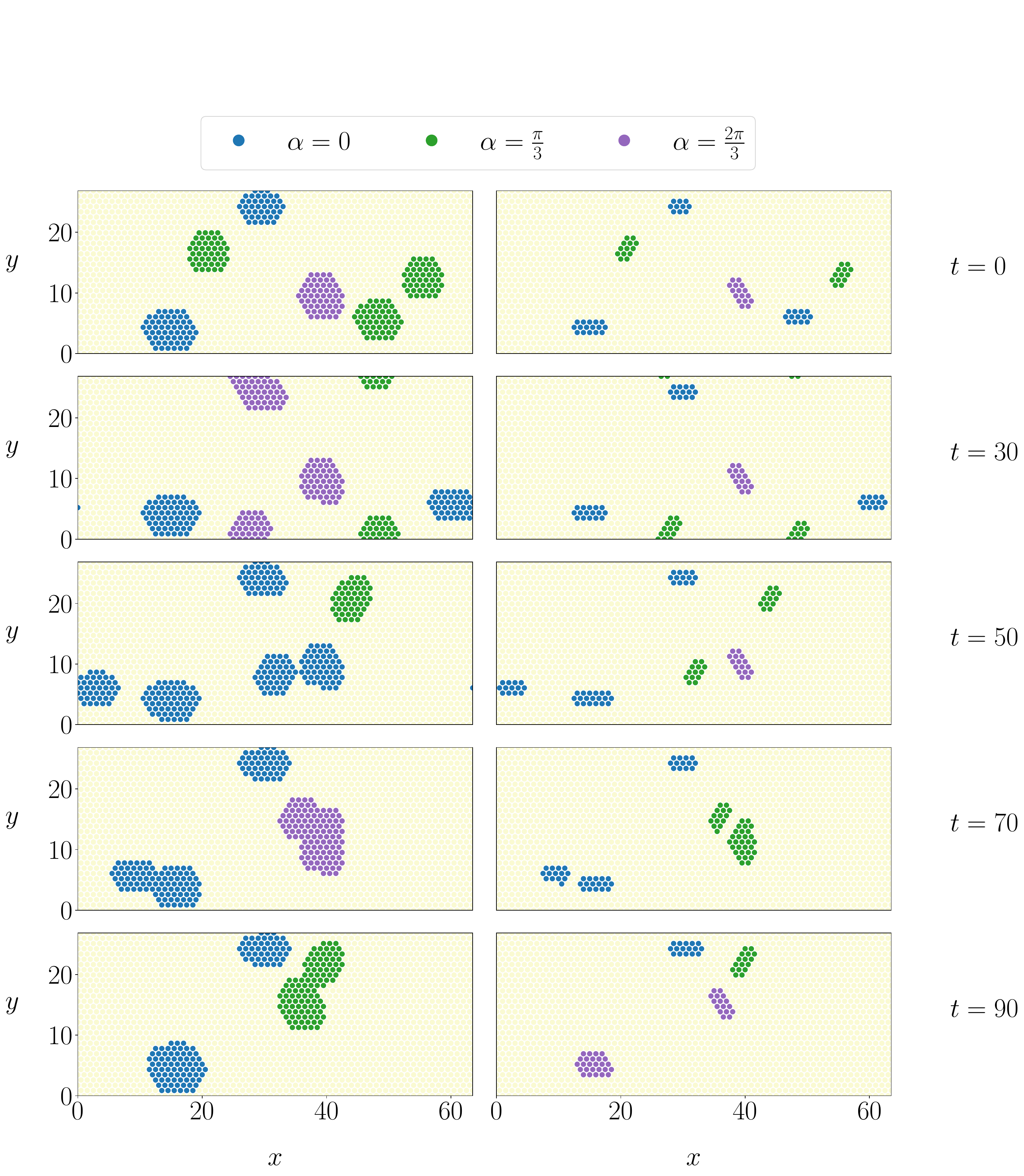}
\caption{Localization regions detected at different times with their estimated directions, using the 2D sliding windows (on the left) and quasi-1D sliding windows with the product approach (on the right), in a simulation of three stationary and three traveling DBs. The localization regions of directions corresponding to $\alpha=0$, $\alpha=\pi/3$, and $\alpha=2\pi/3$ are shown in blue, green, and purple, respectively.}\label{fig:interacions}
\end{figure}

Three stationary and three traveling DBs are simulated till $T_{end}=100$ in the crystal lattice with $N_x=64$ and $N_y=32$ in the following order (from left to right, see $t=0$ in Fig.~\ref{fig:interacions}):
\begin{itemize}
\item stationary DB with $\gamma=0.61$ and $\alpha=0$;
\item traveling DB with $\gamma=0.56$ and $\alpha=\pi/3$;
\item stationary DB with $\gamma=0.41$ and $\alpha=0$;
\item stationary DB with $\gamma=0.52$ and $\alpha=2\pi/3$;
\item traveling DB with $\gamma=0.56$ and $\alpha=0$;
\item traveling DB with $\gamma=-0.54$ and $\alpha=\pi/3$.
\end{itemize}
To identify the localized wave regions and estimate their directionality, the 2D and quasi-1D sliding windows of size $N_d=2$ are used with the product approach. The localization regions together with their directionality at the initial moment $t=0$ as well as at $t=30$, $t=50$, $t=70$, and $t=90$ are shown in Fig.~\ref{fig:interacions}.

As can be seen in Fig.~\ref{fig:interacions}, in all the considered time steps, the localization regions are detected. At $t=0$, the regions of all six DBs are identified and separate independently of the applied sliding window shape. In the case of the quasi-1D sliding windows, the directionality of all DBs is correctly identified, and the detected regions of localizations indeed reflect the quasi-1D nature of DBs. In contrast, in the case of the 2D sliding window, the directionality of one of the localized waves is incorrectly identified. At $t=30$, in the case of the quasi-1D window approach, all the localization regions still preserve the DB quasi-one dimensionality. They are still separate, and their directionality is correctly detected. In the case of the 2D window, the localization regions of two of the waves form one big region, making it impossible to identify their directions. Other wave directions are identified correctly. A similar behavior is observed at $t=50$. In the case of the quasi-1D windows, all regions are separate, and the directions are correctly detected, while using the 2D sliding window, two of the regions are combined again. The localization regions in the quasi-1D sliding window case, at $t=50$, are also geometrically more consistent than in the case of the 2D sliding window. The localized wave interactions have become significant by $t=70$. When the interactions are strong, the geometric and other properties of the interacting waves change. In the case of the quasi-1D sliding windows, two of the waves, at $t=70$, form one localization region because of their collision, while other localization regions are still separate. In the case of the 2D sliding window, only three localization regions are found in the considered time step. The directionality of the DBs whose identified localization regions are separate is correctly detected in both cases. By $t=90$, the collision observed at $t=70$ has ended and, as a result, only one DB out of two participating in this collision has remained. At the same time, a collision of two other DBs occurs. In the quasi-1D sliding window case, at $t=90$, three DB regions together with their directions as well as the localization region corresponding to the collision are correctly identified. In the case of the 2D sliding window, two of the regions corresponding to the localized waves which do not collide form one big region. The region of the remaining wave is separate, and its directionality is correctly detected.

It is important to note that the directionality of a DB collision region is physically undefined because collision regions may not be quasi-1D. Nevertheless, the proposed directionality detection algorithm is able to compute the coefficients~\eqref{eq:correspondence_coefficients} and make a conclusion even about the directionality of a localization region that is not quasi-1D. Moreover, regions of DB collisions cannot be considered DBs themselves; however, the classifiers, described in Section~\ref{sec:classification}, which are applied together with the sliding window method, are capable of detecting the corresponding localization regions.

It can be concluded that, despite the quasi-1D sliding window approach being computationally more expensive, the quasi-1D sliding windows provide better resolution, allowing for distinguishing localized wave regions more accurately, whereas the use of the 2D sliding windows often leads to an inability to distinguish the regions and, as a result, detect the directionality of two localized waves which are located close to each other but still separate. Moreover, the localization regions obtained using the quasi-1D window approach reflect the geometric properties of DBs more precisely in comparison to those obtained with the 2D sliding window approach.

After performing the numerical simulation of six DB interactions, two DB collisions are examined in more detail. Generally, considering DB collisions, different scenarios can occur. For instance, only one DB may remain after a collision, as was seen in the six DB simulation (see Fig.~\ref{fig:interacions}). It is also possible to obtain DBs of a different type or different directionality compared to the generated DBs, as was demonstrated in~\cite{bajars2021two}.

\begin{figure}[t]
\centering
\includegraphics[width=0.73\linewidth]{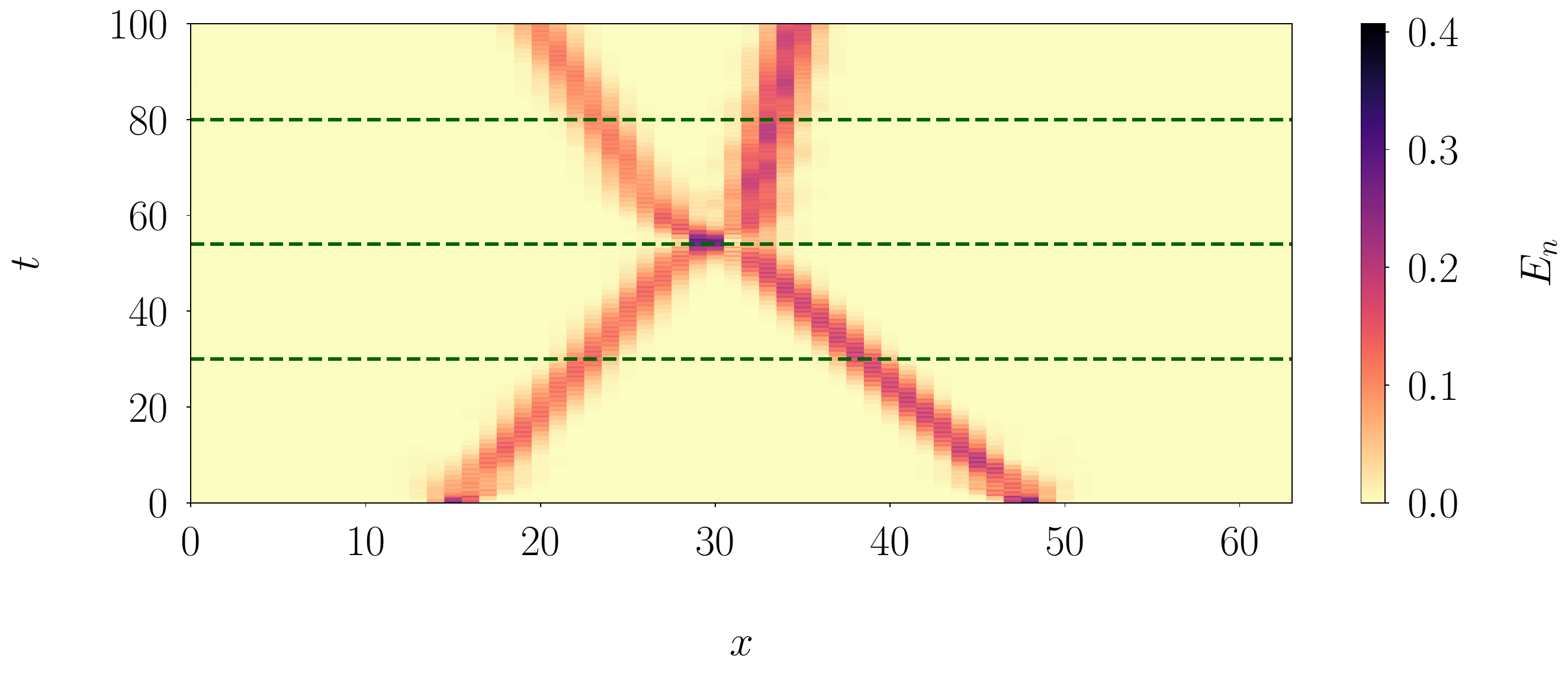}
\caption{Energy density values $E_n$ in the main particle layer of DBs performing a simulation of two traveling DB head-to-head collision.}\label{fig:col_1_energy_density}
\end{figure}

\begin{figure}[t]
\centering
\includegraphics[width=0.75\linewidth]{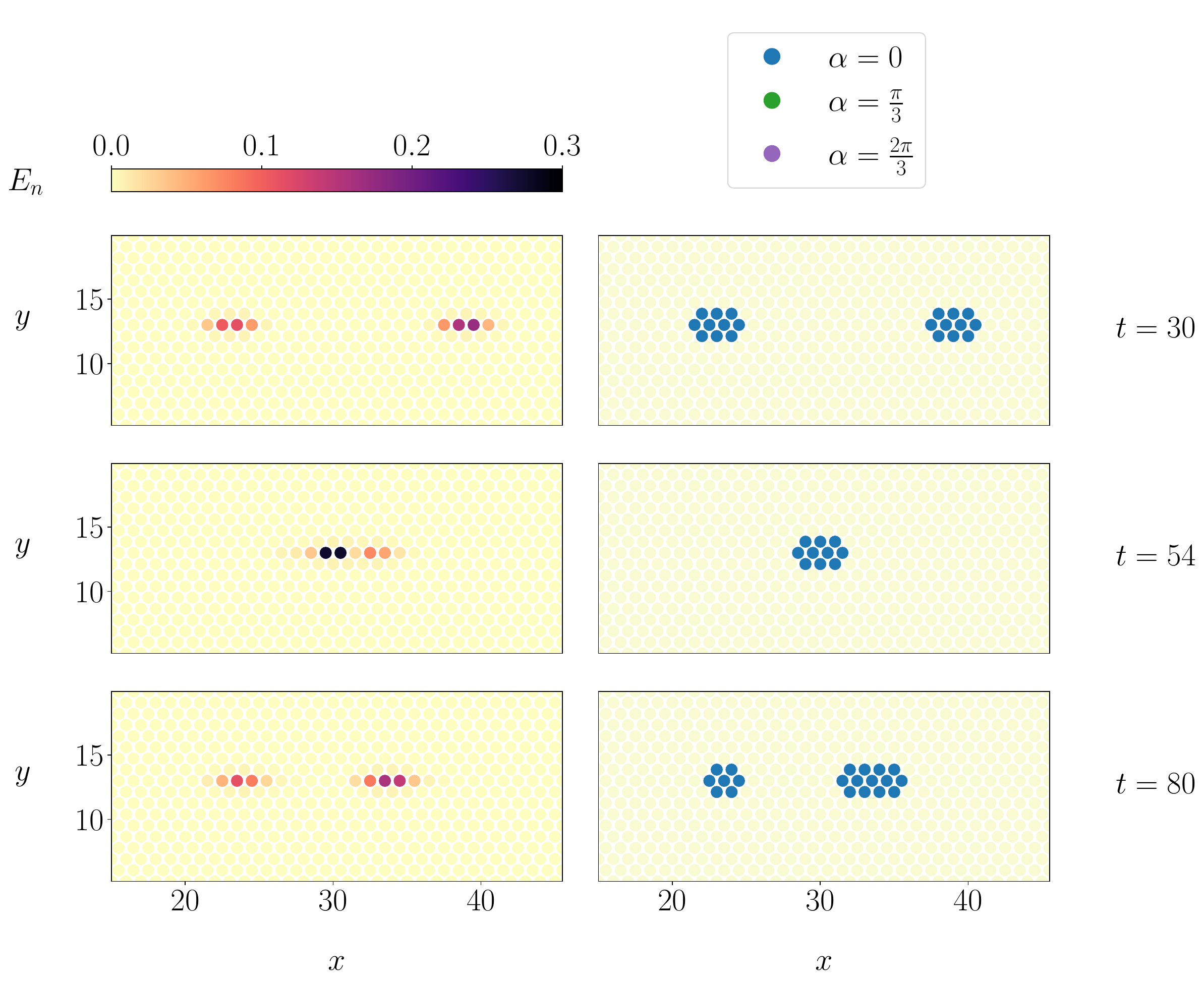}
\caption{Energy density values $E_n$ at different moments in time (on the left) and the detected localization regions with their estimated directions (on the right) in a simulation of two traveling DB head-to-head collision. The localization regions of directions corresponding to $\alpha=0$, $\alpha=\pi/3$, and $\alpha=2\pi/3$ are shown in blue, green, and purple, respectively.}\label{fig:col_1_localization_regions}
\end{figure}

Firstly, a DB head-to-head collision is considered, after which two traveling DBs with the same directionality as the generated waves are obtained. Two traveling DBs with $\alpha=0$ are generated on the opposite sides of the same layer of particles in the crystal lattice with $N_x=64$ and $N_y=32$. The DBs are initialized with $\gamma_l=0.402$ and $\gamma_r=-0.451$, respectively, and the simulation is performed till $T_{end}=100$. Thus, one of the DBs (generated with $\gamma_l$) moves from left to right of the lattice, and another DB (generated with $\gamma_r$) moves from right to left of the lattice. In each simulation time step, the localized wave regions are detected using the quasi-1D sliding windows of size $N_d=2$ with the product approach. The energy density values of the particles in the considered layer in each simulation time step are shown in Fig.~\ref{fig:col_1_energy_density}, while the energy density values of all lattice particles and the detected localization regions at $t=30$, $t=54$, and $t=80$, i.e., before, during, and after the collision, can be seen in Fig.~\ref{fig:col_1_localization_regions}. The moments in time considered in Fig.~\ref{fig:col_1_localization_regions} are shown in Fig.~\ref{fig:col_1_energy_density} with green dashed lines.

It can be seen in Fig.~\ref{fig:col_1_energy_density}~and~\ref{fig:col_1_localization_regions} that, as a result of the collision, two traveling DBs with $\alpha=0$ were formed whose energy is still mostly concentrated on the layer of crystal lattice particles where the initial DBs were generated. Moreover, these DBs move in the same directions as the generated DBs, i.e., one DB moves from left to right, and another one moves from right to left of the lattice. As can be seen in Fig.~\ref{fig:col_1_energy_density}, eventually, each of the obtained DBs propagates with a lower speed compared to the original wave moving in the same direction. The DB propagating from right to left of the lattice is faster than the DB propagating from left to right of the lattice.

As can be seen in Fig.~\ref{fig:col_1_localization_regions}, in each of the considered time steps, all localization regions are detected, preserving the quasi-one-dimensionality of the DBs whenever relevant. Moreover, the directionality of the DBs before and after the collision is also correctly identified. It is important to note that the length of the detected region of localization does not reflect the speed of the corresponding traveling DB, and the detected regions' shapes fluctuate as time progresses, as the lattice particles do the same. It is confirmed by the localization regions detected at $t=80$: the region of localization corresponding to the faster DB is shorter than the one corresponding to the slower DB.

\begin{figure}[t]
\centering
\includegraphics[width=0.6\linewidth]{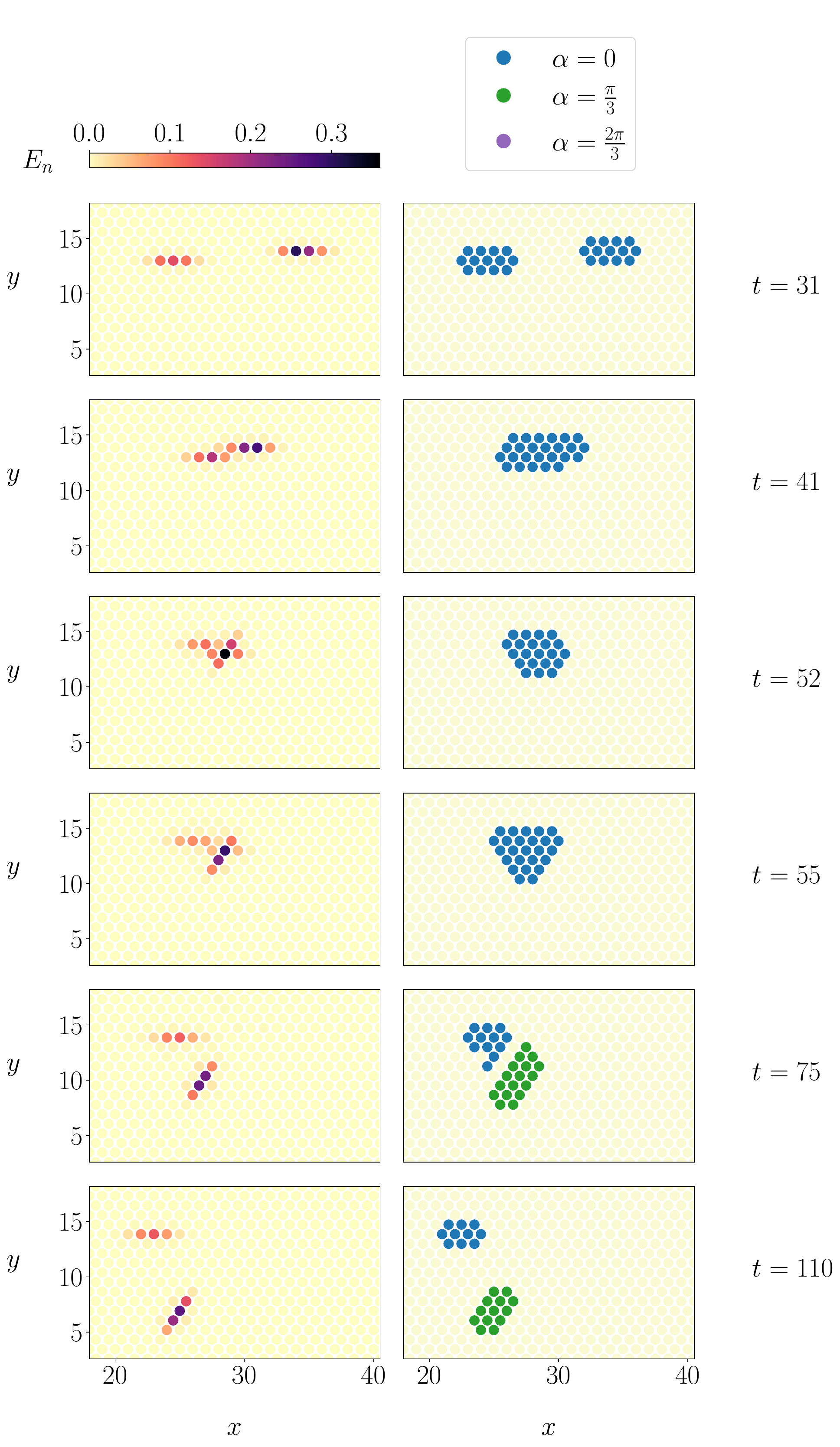}
\caption{Energy density values $E_n$ at different moments in time (on the left) and the detected localization regions with their estimated directions (on the right) in a simulation of two traveling DB head-to-adjacent-head collision. The localization regions of directions corresponding to $\alpha=0$, $\alpha=\pi/3$, and $\alpha=2\pi/3$ are shown in blue, green, and purple, respectively.}\label{fig:col_2_localization_regions}
\end{figure}

A simulation of two equally directed DB collision is further investigated, where two DBs are obtained after the collision, and one of them has a different direction compared to the generated localized waves (see Fig.~\ref{fig:col_2_localization_regions}). As in the previously described simulation, initially, two traveling DBs with $\alpha=0$ are generated on the opposite sides of the crystal lattice with $N_x=64$ and $N_y=32$. However, in this case, the DBs are initialized on the adjacent layers of the crystal lattice particles with $\gamma_l=0.431$ and $\gamma_r=-0.55$, thus, also moving from left to right and from right to left of the lattice, respectively. Hence, it is said that a \emph{head-to-adjacent-head} collision of two traveling DBs is simulated. The simulation is performed till $T_{end}=200$. Similarly to the case of DB head-to-head collision, the quasi-1D sliding windows of size $N_d=2$ with the product approach are used to identify the regions of localization in each simulation time step.

The energy density of all lattice particles as well as the detected regions of localization at $t=31$, $t=41$, $t=52$, $t=55$, $t=75$ and $t=110$ are shown in Fig.~\ref{fig:col_2_localization_regions}. As can be seen, after the collision, two traveling DBs formed: one traveling DB with $\alpha=0$ propagating from right to left of the lattice and another traveling DB with $\alpha=\pi/3$ propagating from top to bottom of the lattice. Moreover, the energy of the DB with $\alpha=0$ is mostly concentrated on the layer of particles where the initial DB traveling from right to left of the lattice was generated.

At $t=31$, the collision has not happened yet. It is seen that the regions of localization are correctly identified and reflect the DB geometric properties. By $t=41$, the collision has started. At this moment, a non-quasi-1D collision region is observed, which is accurately captured by the detected localization region. At $t=52$ and $t=55$, the collision continues, and the energy is redistributed in the collision region, which leads to the constant change of the corresponding detected localization region shape. As previously mentioned, the directionality of the collision region is physically undefined, and, therefore, the correctness of the directions assigned to the corresponding detected localization regions in different time steps cannot be evaluated. By $t=75$, the collision has ended; however, the newly obtained DBs with $\alpha=0$ and $\alpha=\pi/3$, respectively, still strongly interact, which affects the shapes of the detected localization regions. Nevertheless, at $t=75$, the detected regions of localization are already separate, and the directionality of the DBs is correctly identified. At $t=110$, the DB interaction is weaker than at $t=75$, which is reflected by the quasi-1D shape of the detected localization regions. It is also seen that the obtained DBs indeed propagate from right to left and from top to bottom of the lattice, respectively.

It can be concluded that the proposed algorithms for detecting localized wave regions and identifying their directionality in different time steps of a crystal lattice simulation can be effectively applied to perform a numerical study of DB collisions. The obtained localization regions when using the quasi-1D window product approach reflect the geometric properties of DBs and their collision regions with high accuracy.

\FloatBarrier

\section{Conclusions}\label{sec:conclusions}

In this work, it was shown that the data-driven methods presented in~\cite{bajars2022data} for localized wave region detection in 1D crystal lattice simulations can be successfully applied also in the case of the 2D model described in Section~\ref{sec:mathematical_model}. Moreover, highly accurate algorithms were proposed for localized wave directionality identification based on the detected regions of localization. As in the 1D case, the presented methods rely only on locally sampled crystal lattice data in a single simulation time step, which makes them flexible and relatively computationally efficient.

Three types of data collection regions were described in Section~\ref{sec:data_collection}, i.e., the 2D, 1D, and quasi-1D regions, and it was shown in Section~\ref{sec:classification} that, regardless of the choice of the data collection region, highly accurate SVCs can be trained to distinguish between DB and phonon wave data. Moreover, in all the considered cases, the precision and recall values exceed $0.99$. Before training the classifiers, as in \cite{bajars2022data}, the dimensionality of the lattice wave data was reduced. In this work, PCA was applied for this purpose, which allows for reducing the original dataset dimension number even by up to approximately $97\%$, while preserving $95\%$ of the original dataset variance.

In Section~\ref{sec:numerical_results}, the previously trained classifiers were applied in crystal lattice simulations, and different approaches for localized wave region detection were analyzed. While the 2D window approach is computationally the cheapest, using the quasi-1D sliding windows together with the product approach for aggregating the predictions leads to obtaining regions of localization which reflect the quasi-one dimensionality of DBs.

The algorithms for localization region segmentation and directionality detection were also proposed in Section~\ref{sec:numerical_results}. Moreover, the accuracy of the considered method for directionality detection was analyzed, and it was shown that highly precise directionality estimates with accuracy that exceeds $0.99$ can be obtained using all the considered types of sliding windows with both prediction aggregation approaches, when the aggregation is required, except for the 1D sliding windows together with the product approach.

At the end of Section~\ref{sec:numerical_results}, a simulation study was performed to evaluate the effectiveness of the proposed methods when simulating many DBs of different types and directions that propagate and interact. Firstly, a simulation of six DBs was performed. For localization region detection and directionality estimation, the 2D sliding window and the quasi-1D sliding windows with the product approach were applied, and it was shown that using the latter approach allows for obtaining smaller localization regions that better reflect the geometric properties of DBs. The resolution of the quasi-1D sliding window method with the product approach is higher, which allows for distinguishing the regions of localization more accurately. Moreover, this method allows for detecting the DB directionality more precisely. Then, two traveling DB collisions were investigated more closely. In particular, numerical simulations of DB head-to-head and head-to-adjacent-head collisions were performed. In the first case, after the collision, two traveling DBs formed with the same directionality as the initial localized waves, while in the latter case, two traveling DBs were formed, one of which had a different direction compared to the initial localized waves. In each simulation time step, the localization regions were detected using the quasi-1D sliding windows together with the product approach, and the directionality of the corresponding DBs was identified whenever relevant. It was shown that the proposed algorithms for detecting the localized wave regions and their directionality based on locally sampled data can be successfully applied to numerically study collisions of DBs in a 2D hexagonal crystal lattice model. These algorithms provide insights about the geometrical properties of DBs and their collision regions.

The problem of crystal lattice wave multiclass classification, allowing for distinguishing not only between DBs and phonon waves but also between stationary and traveling DBs, was not addressed in this work. It is also important to consider the problem of localization region detection in a thermalized crystal lattice. These challenges, together with data-driven identification of other localized wave properties, are a part of our future work.

\appendix

\section{Training the SVM classifiers}\label{ap:SVC_training}
\renewcommand{\theequation}{A.\arabic{equation}}
\setcounter{equation}{0}

It is considered that $\mathbf{Z}\in\mathbb{R}^{N_{sim}\times d}$ is a low-dimensional representation of one of the lattice wave datasets from Section~\ref{sec:data_collection}, which was obtained by applying PCA. Additionally, a label vector $\mathbf{y}=(y_1,\ldots,\,y_{N_{sim}})^T$ is defined, where $y_i\in\{-1,\,1\}$ is the label corresponding to the $i$-th data instance $\mathbf{z}^i$, contained in the $i$-th row of the low-dimensional data matrix $\mathbf{Z}$, $i=1,\ldots,\,N_{sim}$. If $\mathbf{z}^i$ is an instance of DB data, then $y_i=1$; otherwise, $y_i=-1$. Hence, SVC is a supervised machine learning algorithm.

The main goal of SVC is to find a maximal margin between points that belong to different classes. The decision boundary
\begin{equation}\label{eq:decision_boundary}
\mathbf{w}^T\bm{\phi}(\mathbf{z})+b=0,\quad\mathbf{z}\in\mathbb{R}^d,
\end{equation}
is constructed, where $\mathbf{w}\in\mathbb{R}^D$ is a weight vector, $\bm{\phi}\colon\mathbb{R}^d\to\mathbb{R}^D$, $b\in\mathbb{R}$ and $D\geq d$. It is located in the middle of the margin, and the objective is to maximize the distance from the decision boundary to the nearest data points of each class. The function $\bm{\phi}$ maps data points to the space $\mathbb{R}^D$ where data points of different classes can be linearly separated. In this case, the decision boundary is a hyperplane. If $\bm{\phi}(\mathbf{z})=\mathbf{z}$, for all $\mathbf{z}\in\mathbb{R}^d$, then SVC is called \emph{linear}. The class predicted for a data instance $\mathbf{z}^0$ is $\sgn{(\mathbf{w}^T\bm{\phi}(\mathbf{z}^0)+b)}$. Appropriately scaling $\mathbf{w}$ and $b$, the following optimization problem is solved~\cite{bishop2006pattern}:
\begin{equation}\label{eq:SVC_optimization_problem}
\min_{\mathbf{w},\,\xi,\,b}\left(\frac{1}{2}\mathbf{w}^T\mathbf{w}+\lambda\sum_{i=1}^{N_{sim}}\xi_i\right)
\end{equation}
subject to
\begin{equation}\label{eq:SVC_optimization_constraints}
\begin{aligned}
& y_i(\mathbf{w}^T\bm{\phi}(\mathbf{z}^i)+b)\geq 1-\xi_i, \\
& \xi_i\geq 0,
\end{aligned}
\end{equation}
for all $i=1,\ldots,\,N_{sim}$. The constant $\lambda\geq 0$ is the regularization hyperparameter that determines the extent to which the data points can violate the margin. The nonnegative variables $\xi_i$ are called \emph{slack variables} and defined by
\[
\xi_i=\max(0,\,1-y_i(\mathbf{w}^T\bm{\phi}(\mathbf{z}^i)+b)).
\]
It holds that,
\begin{itemize}
\item if $\xi_i=0$, then the data point $\mathbf{z}^i$ is correctly classified;
\item if $0<\xi_i<1$, then the data point $\mathbf{z}^i$ is inside the margin on the correct side of the decision boundary~\eqref{eq:decision_boundary};
\item if $\xi_i=1$, then the data point $\mathbf{z}^i$ lies on the decision boundary~\eqref{eq:decision_boundary};
\item if $\xi_i>1$, then the data point $\mathbf{z}^i$ is incorrectly classified.
\end{itemize}

The optimization problem~\eqref{eq:SVC_optimization_problem}--\eqref{eq:SVC_optimization_constraints} is a convex programming problem. Introducing Lagrange multipliers $a_i\geq 0$ and $\mu_i\geq 0$ for the constraints~\eqref{eq:SVC_optimization_constraints}, $i=1,\ldots,\,N_{sim}$, and taking into account Karush-Kuhn-Tucker conditions, the corresponding Lagrangian function is
\[
L(\mathbf{a})=\sum_{i=1}^{N_{sim}}a_i-\frac{1}{2}\sum_{i=1}^{N_{sim}}\sum_{j=1}^{N_{sim}}a_ia_jy_iy_j\mathcal{K}(\mathbf{z}^i,\,\mathbf{z}^j),
\]
where $\mathbf{a}=(a_1,\ldots,\,a_{N_{sim}})^T$ and
\[
\mathcal{K}(\mathbf{z}^i,\,\mathbf{z}^j)=\bm{\phi}(\mathbf{z}^i)^T\bm{\phi}(\mathbf{z}^j)
\]
is a symmetric kernel function. It can be defined without explicit knowledge of $\bm{\phi}$. To solve the optimization problem~\eqref{eq:SVC_optimization_problem}-\eqref{eq:SVC_optimization_constraints}, its dual quadratic programming problem is solved together with some additional constraints derived from Karush-Kuhn-Tucker conditions~\cite{bishop2006pattern, chang2011libsvm}:
\begin{equation}\label{eq:SVC_dual_optimization_problem}
\max_{\mathbf{a}}{L(\mathbf{a})}
\end{equation}
subject to
\begin{equation}\label{eq:SVC_dual_optimization_constraints}
\begin{aligned}
& 0\leq a_i\leq \lambda,\quad\text{for all}\quad i=1,\ldots,\,N_{sim}, \\
& \sum_{i=1}^{N_{sim}} a_iy_i=0.
\end{aligned}
\end{equation}
After solving the optimization problem~\eqref{eq:SVC_dual_optimization_problem}-\eqref{eq:SVC_dual_optimization_constraints}, the optimal $\mathbf{w}$ is computed as follows:
\[
\mathbf{w}=\sum_{i=1}^{N_{sim}}a_iy_i\bm{\phi}(\mathbf{z}^i).
\]
If $a_i=0$, then the data point $\mathbf{z}^i$ does not impact the prediction. Moreover, it can be derived that, if $0<a_i<\lambda$, then $\xi_i=0$. Therefore, for all $i\in\mathcal{M}=\{i\in\{1,\ldots,\,N_{sim}\}\mid 0<a_i<\lambda\}$, it holds that
\begin{equation}\label{eq:SVC_deriving_b}
y_i\sgn{(\mathbf{w}^T\bm{\phi}(\mathbf{z}^i)+b)}=y_i\left(\sum_{\substack{j=1 \\ a_j>0}}^{N_{sim}} a_jy_j\mathcal{K}(\mathbf{z}^i,\,\mathbf{z}^j)+b\right)=1.
\end{equation}
Multiplying both sides of~\eqref{eq:SVC_deriving_b} by $y_i$ and averaging over $i\in\mathcal{M}$, a numerically stable solution for $b$ is obtained:
\[
b=\frac{1}{|\mathcal{M}|}\sum_{i\in\mathcal{M}}\left(y_i-\sum_{\substack{\substack{j=1 \\ a_j>0}}}^{N_{sim}} a_jy_j\mathcal{K}(\mathbf{z}^i,\,\mathbf{z}^j)\right).
\]

\section*{Acknowledgements}

F.~Kozirevs and J.~Bajārs acknowledge financial support from the Latvian Council of Science, project No.~lzp-2024/1-0207 ``Development of structure- and data-driven methods for analysis and control of complex dynamical systems''.

\bibliographystyle{unsrt}  
\bibliography{mybibfile}  

\end{document}